\documentclass[12pt]{article}
\usepackage{natbib,amssymb,amsmath}
\usepackage{amsfonts}
\usepackage{graphicx}
\usepackage{amsmath}
\usepackage{amssymb}
\usepackage{color}

\newcommand{\be}{\begin{equation}}
\newcommand{\ee}{\end{equation}}

\begin{document}

\begin{center}
{\large Amplification of wave groups in the forced nonlinear Schr\"odinger equation}
\end{center}

\begin{center}

Montri Maleewong$^{1}$ and Roger H.J. Grimshaw$^2$  \\ \vspace{0.3cm}

$^1$ Department of Mathematics, Faculty of Science, Kasetsart University, \\ Bangkok, 10900, Thailand \\
$^2$ Department of Mathematics,  University College London, \\ London WC1E 6BT, UK \\

\end{center}

\begin{abstract}
In many physical contexts, notably including deep water waves, 
modulation instability in one space dimension is often studied using  the nonlinear
Schr\"odinger equation.  The principal solutions of interest  are solitons and breathers 
which are adopted as models  of wave packets.  The Peregrine breather in particular is often invoked as a model of a rogue wave.  In this paper we add a linear growth term 
to the nonlinear Schr\"odinger equation to model the amplification of propagating wave groups.
This is motivated by an application to wind-generated water waves, but this forced 
nonlinear  Schr\"odinger equation has potentially much wider applicability.
We describe a series of numerical simulations  which in the absence of the forcing term would
generate solitons and/or  breathers.  We find that overall the effect of the forcing term is 
to favour the generation of solitons  with  amplitudes growing at twice the linear growth rate
over the generation of breathers. 

\end{abstract}

\section{Introduction}

It is well known that modulation instability, that is, 
the exponential growth of long wave perturbations to a periodic plane wave,  leads to the
formation of nonlinear wave packets, and sometimes to  rogue waves.
This process is often modelled by the nonlinear Schr\"odinger equation (NLS), 
and then  the nonlinear wave packets can be represented by the soliton and breather  solutions of the NLS, 
while  rogue waves are often modelled  by  the Peregrine breather, see for instance 
\citet{KH09, OS10, gr13} in the deep water wave context.  The process occurs in 
many other physical contexts, see \citet{gr10, ccg19} for internal wave applications, 
and the related articles in that special issue for other cases.
In this paper we develop the formulation in the water wave context to be specific but the outcome can be 
applied to many other physical contexts. 

Wave packets in one horizontal space dimension are given by  
\be\label{packet}
\zeta \,=\, \delta A(X, T) \exp{(i\theta )} + \hbox{c.c.} \, + \cdots \,,
\ee
\be\label{scale}
\hbox{where} \quad \theta =kx -\omega (k)t \,, \, X=\delta (x-c_g t) \,, \, T=\delta^2 t \,.
\ee
\be\label{disp}
\omega^2 (k) = \frac{g}{h}q\sigma  \,, \quad 
c_g = \omega_k =\frac{\omega}{2k}\{1 + \frac{q}{\sigma}(1-\sigma^2 )\}  \,, 
\quad q  = kh \,,   \quad  \sigma = \tanh{q}\,. 
\ee
Here $\zeta (x,t)$ is the water surface elevation above the undisturbed depth $h$, and 
$k$ is the carrier wavenumber, while the wave  frequency $\omega (k)$ satisfies the 
linear dispersion relation. $A(x,t)$ is the slowly varying  wave amplitude, and at leading order
 the wave packet moves with the group velocity  $c_g = \omega_k $.  $\delta $, $0 < \delta \ll 1$ is a small dimensionless parameter 
 measuring the wave amplitude and dispersion about the dominant wavenumber $k$. 
 The leading order omitted terms in (\ref{packet}) are  $O(\delta^2 )$ second harmonic and mean flow terms.\\
 
A multi-scale asymptotic expansion in $\delta $ in which the  linear dispersive effects are 
scaled to balance the leading order nonlinear effects leads to the NLS equation,
see \citet{bn67, zh68, ho72} and the review by \citet{gr07}.
\be\label{nls0}
iA_T + \lambda A_{XX} + \mu |A|^2 A\,= 0 \,, \quad 
\lambda = \frac{c_{gk}}{2} \,.
\ee
The coefficient $\mu $ of the nonlinear term is given by
\be\label{mu}
\mu = -\frac{k^{2}\omega }{4  \sigma^4 } (9\sigma^4 -10\sigma^2 +9)
+\frac{\omega^3 }{2\sigma^3 (gh -c_{g}^{2})}
(2\sigma (3-\sigma^2 ) +3q(1-\sigma^2 )^2 )\,.
\ee
In deep water ($q \to \infty$) the second term vanishes, and the
coefficient $\mu \to -2\omega k^2 <  0$. 
In general $\mu < 0( > 0)$ according as $q > q_c ( q< q_c)$, 
where $q_c = 1.363$.  Modulation instability occurs when $\mu \lambda > 0$.
For water waves $\lambda < 0$ and so modulation instability occurs for 
waves in deep water when  $\mu < 0$, $q > q_c $.
Similar expansions apply in many other physical systems, again leading to the 
NLS equation (\ref{nls0}).   The main difference is the linear dispersion
relation (\ref{disp}) and in the expressions for the coefficients $\lambda, \mu$,
see \citet{gr07, ap10}  for instance.  We note that in this water wave context the 
wave amplitude from (\ref{packet})  is $2 \delta A$ and is required to be small since 
$\delta \ll 1$ but $\delta $ itself  does not appear explicitly in (\ref{nls0}).

In this paper we are concerned with the effect of  forcing on modulation instability.
We model this by  extending the NLS equation (\ref{nls0}) to a forced NLS equation (fNLS)
by the addition of a linear forcing term, see for instance 
\citet{le07, tkpg08, mkmk13, br14, sl15, gr18, gr19a, gr19b} in the wind wave context
\be\label{nls}
iA_T + \lambda A_{XX} + \mu |A|^2 A\,= i\Delta A \,. \quad 
\ee
The forcing is modelled by the  linear growth rate term with coefficient $\Delta  >0$.
Various expressions can be found in the literature, the most well-known being that originally derived 
 by \citet{mi57} and subsequently adapted and  modified in various ways, see for instance
 \citet{mi93, ms93, ja04, saj06, shd14, zrp17, gr18}. Here our concern  is
 with the effect of $\Delta$ on modulation instability   and wave packet, or breather, formation. 
 The effect of forcing on modulation instability has been examined in the 
 present one space-dimension framework for deep water waves by
\citet{le07, tkpg08, br14, sl15, gr18}. Here we extend these studies which were mostly
concerned with the evolution of wave spectra, by focusing on the development 
of wave packets  through comprehensive numerical simulations of the
 fNLS equation ({\ref{nls}). The formulation of the problem is presented in section 2.
 In section 3 we present these numerical simulations and some accompanying analysis.
 We conclude in section 4.

\section{Formulation}

We consider the case when there is modulation instability, so that 
$\mu < 0, \,\lambda<0$ ($q  > q_c $). Then fNLS (\ref{nls}) can be expressed in  canonical form 
\be\label{fnls}
    i \epsilon Q_T + \epsilon^2 Q_{XX} + 2 |Q|^2 Q \ = \ i \Delta Q \,.
\ee
This canonical form is  achieved through the change of variables
\be\label{transform}
Q = \{\frac{|\mu|}{2}\}^{1/2}\bar{A}, \quad \tilde{X} = \frac{\epsilon}{|\lambda|^{1/2}}X, 
\quad \tilde{T} = \epsilon T, 
\ee
Here  we have introduced the free parameter $\epsilon $ as it is   useful  to represent 
the scaling properties of the NLS equation.
In the small $\epsilon$ limit an asymptotic procedure can be used to describe the 
generation of a  family of Peregrine breathers from a modulated plane periodic wave, see 
\cite{gr13} for an application to water waves.

The fNLS equation (\ref{fnls}) has the energy law
 
\be\label{energy}
E(T) = E(0) \exp{(2\Delta T/\epsilon )} \,, \quad E(T) = \int^{X_L}_{-X_L} |Q(T,X)|^2 dX \,.
\ee
Here if $X=\pm X_L$ in an infinite domain then $Q(X,T)$ must decay sufficiently fast at infinity, 
otherwise in a finite  domain periodic boundary conditions  are applied at $X_L$.
The expression (\ref{energy}) can be used to estimate the growth of the wave amplitude  
as explained in our previous work, see \citet{ghj18}.   
Briefly if in the absence of forcing the solution is $Q(X,T: M)$ 
where $M$ is a free amplitude parameter, then substitution  into 
(\ref{energy}) yields an estimate for the growth $M$ under forcing. 
This is used here as a guide to interpreting each of  the cases we consider. 
In the absence of forcing modulation instability can be measured by the 
Benjamin-Feir index ($\hbox{BFI}$),  the ratio of wave steepness (nonlinearity) to spectral bandwidth (dispersion)
and in the  absence of forcing $\hbox{BFI} \approx 1/\epsilon$, see \citet{gr13}. Using the change of variables 
$Q = \tilde{Q}\exp{(\Delta T/\epsilon )}$ it is readily shown that in the presence of forcing this becomes 
$\exp{(\Delta T/\epsilon )}/\epsilon  $.

The forced NLS equation (\ref{fnls}) is solved numerically on the periodic domain $-L<x<L$, 
using a Fourier spectral method in space and a  Runge-Kutta approximation in time. 
We set $L=30-150$ to minimise boundary truncation effects. 
With this periodic boundary condition, we choose modulation scales so that the solutions decay 
to the initial background at both ends of the domain well within numerical error.  
 In most cases of the numerical simulations, 
we set the number of mesh points as $4096$ and $dT=5e-06$  that satisfies 
numerical stability condition in the Fourier and time domains.

\section{Numerical simulations}

We consider four cases of initial conditions for the forced NLS equation (\ref{fnls}). Each case represents
the generation of solitons and/or breathers.

\smallskip
\subsection{Case 1}
When $\Delta = 0$, the Peregrine breather is given by, see \citet{pb83,chab16},
\be\label{PB}
Q(X,T) \ = \ M \left[ 1 - \frac{4(1+4i\tau)}{1+ 4 \chi^2 + 16 \tau^2 }  
\right] \exp{(2i\tau)} \,, \quad \chi = \frac {M X}{\epsilon },
\quad \tau = \frac{M^2 T}{\epsilon} \,. 
\ee
When $\Delta \ne 0$ we solve numerically the forced NLS equation (\ref{fnls}) with the initial condition
corresponding to this Peregrine breather (\ref{PB}) at $T= T_0 < 0$. 
With $\epsilon =1, M=1, T_0 = -2, \Delta = 0$ and 0.2,  the results are  shown in figures \ref{PB1}-\ref{PB5}.
In the forced case, the amplitude initially grows exponentially at the rate 
$2\Delta $ as shown in figure \ref{PB5} that agrees with the asymptotic prediction by 
\citet{gr19b} using the energy law (\ref{energy}) as described above. 
Note that $E(T)$ in (\ref{energy}) scales as $M^2 \, M^{-1} \epsilon  = M \, \epsilon$ 
modulo a  dependence on $\tau $,
\be\label{energyPB}
 E_{PB}(T) = M\, \epsilon \,J(\tau) \,,
 \ee
 $$ J (\tau ) =  \int^{\chi_L}_{-\chi_L}  | 1 - \frac{4(1+4i\tau)}{1+ 4 \chi^2 + 16 \tau^2 }|^2   d\chi  \,. $$
 Here $\chi_L$ is chosen sufficiently large so that $|Q|$ in (\ref{PB}) has decayed to $M$ there.  
 The function $J(\tau ) \to 2\chi_L$  as $\tau \to \pm \infty $ and has a maximum value at 
 $\tau =0$ on a time scale where $\tau $ is order unity, so that $T$ is order $\epsilon M^{-2}$, much slower than 
 $ \epsilon \Delta^{-1}$ for our parameter choices. 
 In our simulations the amplitude reaches the first peak around $T=-0.5$ 
instead of $T=0$ as in the unforced case, and  then instead of  subsiding to zero as $T\to \infty $,
exhibits several oscillations of increasing amplitude, which appear to be the generation of successive Peregrine breathers.  The outcome resembles the family of Peregrine breathers  to those described in Case 3 below. \\

\smallskip
\subsection{Case 2}
When $\Delta = 0$ there is an exact soliton solution, see \citet{gr07,chab16}, 
\be\label{SOL}
Q(X, T) = M \hbox{sech}(\Theta )\exp{(i\Phi )} \,, \quad
 \Theta = \Gamma (X- VT)\,, \quad \Phi = \hat{K}X- \Omega T \,, 
 \ee
$$ \hbox{where} \quad \Gamma =  \frac{M}{\epsilon }\,,  \quad V= 4\hat{K}\,, \quad  \Omega = \epsilon^2 \hat{K}^2  - \frac{M^2}{\epsilon^2 }  \,. $$ \\
The evolution of $|Q|$ with $\epsilon = 1, M=2, \hat{K}=-2$ and $\Delta=0$ is shown in 
figure \ref{fig2-1}. The soliton is moving with constant amplitude and speed $V=8$ as predicted.
With  forcing $\Delta =0.2$ the evolution of $|Q|$ with $\epsilon = 1, M=2, \hat{K}=-2$ is shown in figure \ref{fig2-2}. 
The soliton is moving with an exponentially increasing amplitude at the rate of $2\Delta $ as shown in 
 figure \ref{fig-case2-theory}. This agrees with the asymptotic prediction of \citet{gr19a} using the energy law (\ref{energy}), while the  speed is hardly changed.
 In the forced case, the amplitude initially grows exponentially at the rate 
$2\Delta $ as shown in figure \ref{PB5} that agrees with the asymptotic prediction by 
\citet{gr19b} using the energy law (\ref{energy}) as described above as 
here $E(T) = 2 \,M \, \epsilon$.
The amplitude of the soliton grows rapidly after $T>7$ and we infer that  the solution has become 
unstable.  When the forcing is turned off after $T>4$, the amplitude of the moving soliton is constant. \\

\smallskip
\subsection{Case 3}
The initial condition is a slowly-varying long wave perturbation. 
\be\label{PBsic}
Q(X, 0) \ = \ M \,\text{sech}(\gamma X) \,.
\ee
Note  that we only show cases with  $M=1$ as $M$ can be absorbed into the small parameter 
$\delta $ in the derivation of (\ref{fnls}), but we did run some simulations for a larger $M=2$.    
Also  $\gamma $ and $\epsilon $ are not independent parameters, as rescaling 
$\tilde{X} = \gamma X$  is equivalent to replacing $\epsilon $ by $\tilde{\epsilon } = \gamma \epsilon $ and then adjusting the time scale $\tilde{T} = \gamma T$.  
Nevertheless we shall vary both $\epsilon $ and $\gamma $, as well as $\Delta $. 
For small $\epsilon$, the dispersion is initially weak and in the absence of forcing 
the solution evolves into a  gradient catastrophe, followed by the generation of a 
family of  Peregrine breathers, see \citet{gr13}.

\smallskip
\noindent{\textbf{(3a) Without forcing ($\Delta =0$)}.} \\
The outcome for the initial condition (\ref{PBsic}) when $\epsilon = 1/33$, $\gamma=1$,
$M=1$  and $\Delta=0$ are shown in figures \ref{PB-surf} and \ref{PB-gradient}. There is a gradient catastrophe at $T=0.25$ and the generation of a family of Peregrine breathers, in agreement with the theory and numerical predictions described by \citet{gr13}.  A case (not shown here) with $\gamma = 1$, $\epsilon=1/33$
but with a larger $M=2$ was similar.   For this larger of $M=2$ the gradient catastrophe occurs 
earlier around $T=0.1$ and this then generates a family of breathers 
but with less ordered  behaviour.  \\

\smallskip
\noindent{\textbf{(3b) With forcing ($\Delta > 0$)}.} \\
The forced NLS (\ref{fnls}) was solved with the initial condition (\ref{PBsic}) 
for various cases with an initial wavenumber  
$\gamma = 0.5, 1.0, 2.0$, $\epsilon=0.03, 0.1, 0.2, 0.5, 1.0$ and $\Delta = 0.00, 0.06$. 
Note that here $E(0) = 2M^2/\gamma $ in the energy law (\ref{energy}) where 
unlike cases 1 and 2, $M, \gamma$ are independent parameters and so both the 
amplitude $M$ and the wavenumber $\gamma $ can be affected by the forcing. 
We show some representative outcomes here.

\smallskip
\noindent
The case when $\gamma=1$, $\epsilon=1/33$ and $\Delta=0.00, 0.06$ is shown in figure 
\ref{fig-sech-gam1-ep003-a}. Without forcing there is again the initial generation of  breathers 
as shown in figures \ref{PB-surf}-\ref{PB-gradient}, but with the longer time simulation, 
the breathers combine to form some propagating solitons. As the forcing is increased the 
breathers are suppressed  and for $\Delta = 0.06$ many  stationary solitons with 
growing amplitudes form.  

\smallskip
\noindent
The case when  $\gamma=0.5$, $\epsilon=0.1$ and  $\Delta =0.00, 0.06$ is shown in figure
\ref{fig-sech-gam05-ep01}. Here $\tilde{\epsilon} = 0.05$ and is comparable to the case when 
$\gamma =1$, $\epsilon = 1/33$ as the values  of
$\tilde{\epsilon} =0.05$ and $\epsilon = 1/33$ are close.


 \smallskip
 \noindent
The case when   $\gamma=0.5$, $\epsilon=0.2$ and $\Delta=0.00, 0.06$ is  shown in figure 
\ref{fig-sech-gam05-ep02}. Here $\tilde{\epsilon} = 0.1$, but in comparison with the previous case shown in
 figure \ref{fig-sech-gam05-ep01}, without forcing there is no sign of the emergence of solitons, 
 and instead a breather family forms and then re-forms.  
  However, in the forced simulations, solitons emerge as in the previous case in figure
 \ref{fig-sech-gam05-ep01} and as the forcing is increased there is a transition to just a 
 few stationary solitons with growing amplitudes.


\smallskip
\noindent
The case when $\gamma=0.5$, $\epsilon=1.0$ and $\Delta=0.00, 0.06$ is shown in figure 
\ref{fig-sech-gam05-ep1}. Here $\tilde{\epsilon} = 0.5$, but unlike the two  previous cases shown in
figures  \ref{fig-sech-gam05-ep01} and \ref{fig-sech-gam05-ep02} when a periodic chain of breathers forms.
This case is converted to several growing solitons, now in the forced simulations 
 only a single stationary soliton with a growing amplitude emerges.
Plot of the maximum value of $|Q(X,T)|$ for each $T$ and $-L<X<L$ is shown in figure \ref{fig-case3-theory}. The amplitude for this forced NLS case grows oscillatory with  an overall growth rate $2\Delta$.
After $T>15$ the forcing $\Delta=0.06$ is turned off, the maximum amplitude is shown in figure
\ref{fig-case3-theory-turn}; it does not increase exponentially but it changes periodically.


\smallskip
\noindent
The case when $\gamma=2.0$, $\epsilon=0.5$ and $\Delta=0.00, 0.06$ is shown in figure 
\ref{fig-sech-gam2-ep05}. As $\tilde{\epsilon} =1.0$ this is equivalent to Case 2 and indeed only a single 
stationary soliton forms, with a growing amplitude at exactly the predicted exponential rate of 
$2\Delta $, see \citet{gr19a}.

\smallskip
\noindent
The case when $\gamma=2.0$, $\epsilon=1.0$ and $\Delta=0.00, 0.06$ is shown in figure
 \ref{fig-sech-gam2-ep1}. Here $\tilde{\epsilon} =2.0$ is larger and  a different picture emerges, looking 
 more like the modulation instability cases shown in Case 4 below. There is evidence of the formation of 
 both breathers and solitons with amplitudes increasing.
 
\smallskip
\subsection{Case 4}

The initial condition is a long-wave periodic perturbation with wavenumber $K$, 
\be\label{MIic}
Q(X, 0) \ = \ M(1 + \alpha \cos{KX}) \,,
\ee
where $0 < \alpha \ll 1$.   When $\Delta =0$ there is modulation instability for
 $\epsilon K < \sqrt{2}|M|$, and maximum growth when $ \epsilon K = \sqrt{1/2}|M|$. 
 We fix $\alpha =0.1$ and show some representative simulations varying $M$, $K$
 and $\Delta $.   With this  initial condition (\ref{MIic}) where the initial amplitude 
 does not decay at  the boundaries, the computational domain $L$ is set large enough 
 and given by a multiple of $2\pi/K$ to prevent some spurious effects from the boundaries. 
 Since we have applied a periodic boundary condition in the Fourier spectral method, 
 some waves will appear to come from the boundaries in this case. \\

\smallskip	
\noindent{\textbf{(4a) Without forcing ($\Delta = 0$)}.} \\

We  examined the case when $M=0.1$, $\epsilon=1/33$ and $K=0.1$.
This parameter setting is at the long-wave end of the modulation instability regime, 
and the outcome is a family of Peregrine breathers very similar to that shown in Case (3a) 
in figure \ref{PB-surf}.   Cases when  $M=1$, $\epsilon=1$, $K=\sqrt{0.1}$  
and $M=1$, $\epsilon=1$, $K=\sqrt{0.5}$ were also investigated.
Both these cases are within the modulation instability regime, and we found the generation of 
breathers as has been demonstrated in many works, see for instance \citet{OS10}.  
The case $M=1$, $\epsilon=1$,$K=4$ is formally outside the modulation instability regime.  
Nevertheless a  periodic breather chain develops after $T=18$ with amplitudes less than $3$, 
see figure \ref{fig-ep4a}. \\

%

\smallskip
\noindent{\textbf{(4b) With forcing ($\Delta > 0$)}.} \\
The case when $M=0.1$, $\epsilon=1/33$, $K=0.1$ and $\Delta=0.02$ is shown in figure \ref{case4del002}. 
As the forcing is increased the family of Peregrine breathers is converted to many stationary solitons 
with amplitudes increasing in time and with a short length scale. The cases when $M=1.0$, $\epsilon=1.0$, $K=\sqrt{0.1}$ and $\Delta=0.02$ is shown in figure 
\ref{case4del002} (top-right) and the case of $M=1.0$, $\epsilon=1.0$, $K=\sqrt{0.5}$ 
with $\Delta=0.02$  is shown in figure \ref{case4del002} (bottom-left).
Plots of the maximum amplitude versus time $T$ when $K=\sqrt{0.1}$ for various values of 
$\Delta$ is shown in figure \ref{fig-case4-theory2}. The amplitudes grow with the 
exponential growth rate $2\Delta$. 
The case when $M=1.0$, $\epsilon=1.0$, $K=4$ and $\Delta=0.02$
 is shown in figure \ref{case4del002} (bottom-right).
 In all these cases as the forcing is increased the breathers are eliminated 
 and progressively fewer solitons are formed with growing amplitudes.
 Plots of the maximum amplitude versus time $T$ for various values of $\Delta$ is 
 shown in figure \ref{fig-case4-theory1}, the amplitudes grow with the 
 exponential growth rate $2\Delta$. 
The maximum amplitude does not grow at the early time steps, as a certain time interval is required for the excitation of the wave amplitude.
For the unforced case, when the maximum amplitude grows to its maximum average, 
it does not grow further. The higher the value of forcing amplitude, the smaller 
the time interval required for the excitation.
The number and trajectories of these solitons appears to depend quite sensitively on the parameter settings. 
When the forcing is turned off after $T>15\, s$, the contour plot of soliton formed is shown in 
figure \ref{case4-turn} (left). The trains of solitons (large waves) interact with each other. 
Since there is no forcing after $T>15\, s$, the maximum amplitude does not grow. 
It maintains  the mean value which is approximately equal to the final amplitude at $T=15$ 
as shown in figure \ref{case4-turn} (right). 

\subsection{Initial random noise}

In this section, we investigate the effect of initial random noise  on the growth rate of 
the wave amplitude for each case presented in the previous sections.   We impose   
random noise  at the initial time step and  investigate the dynamics of breathers, solitons and modulation instability. 
The initial random noise is set by
\be\label{random}
Q(X,T_0) + M_R \cdot \text{Rand} \,,
\ee
where $Q(X,T_0)$ is the initial condition as above when there is no random noise. 
$\text{Rand}$ is a complex random number where the real and imaginary parts are 
uniformly distributed values ranged over $[0,1]$. Random noise is inserted on 
every grid point for $-20<X<20$ and the  value of $\text{Rand}$ is zero outside this region. 
$M_R$ is the magnitude of the random noise.  
We set $M_R=0.1$ for case 1 and $M_R=0.5$ for cases 2-4. It is approximately 
$25\%$ of the initial amplitude. We  numerically investigate how this initial random noise 
affects  the growth rate under forcing. The  results for each case are shown as follows.\\

\noindent
{\bf Case 1}: The effect of initial random noise on the Peregrine breather is shown in 
figure \ref{case1-rand} with $\Delta=0.2$, $\epsilon=1$ and $M=1$.
It can be compared to the case  without initial noise  shown in figure \ref{PB3}.
The main feature of breathers can still be seen  in figure \ref{case1-rand} (left).
 The maximum amplitude over the entire domain is shown in figure \ref{case1-rand} (right)
 and is comparable with the theoretical growth rate $2\Delta$.  
 The initial random noise again affects  the increment of maximum amplitude 
 earlier than $T=0$, as before when $-0.5 <T$. 
 The position of these large amplitude waves remains relatively stationary.\\

 \noindent
{\bf Case 2}: The evolution of a moving soliton with initial random noise is shown in figure \ref{case2-rand} (left). 
Here $\epsilon=1$, $M=2$, $\hat{K}=-2$ and $\Delta=0.2$. This result can be compared with the case  without initial noise as shown in figure \ref{fig2-2}. The soliton is moving with an exponentially increasing amplitude by the growth rate $2 \Delta$. Plot of the maximum amplitude is shown in figure \ref{case2-rand} (right). 
Unlike case 1, the maximum amplitude grows monotonically. The results  with and without initial 
random noise are similar except that now two stationary solitons are generated  downstream 
while the leading soliton travels with the theoretical speed  upstream. We infer that
initial random noise that has a large enough amplitude can generate some new stationary  
soliton waves.  \\

\noindent
{\bf Case 3}:  In the absence of forcing and initial random noise, a periodic chain of breathers forms. 
The location of the breathers is stationary along $X=0$ as shown in figure \ref{fig-sech-gam05-ep1}. 
The result with initial random noise is shown in figure \ref{case3-rand} (left). 
In this simulation initial random noise shifts the location of the periodic chain of breathers 
slightly to $X < 0$,  with the same of order of amplitude and with an indication of  
a second chain forming in  $X>0$. 
For a forcing case with $\Delta=0.06$, the maximum amplitude is increased, 
see figures \ref{case3-rand} (right),  with the mean growth rate $2 \Delta$, see figure \ref{case3-rand2}. \\



\noindent
{\bf Case 4}: The results with initial random noise are shown in figure \ref{case4-rand} 
for $M=1$, $\epsilon =1$, $K=4$ and $\Delta = 0, 0.04$. 
There is no modulation instability in this case. In the absence of forcing, 
see figure \ref{case4-rand} (left), periodic plane waves are generated. 
Initial random noise perturbs the pattern of these waves while the 
maximum amplitude over the entire domain is preserved. 
In contrast for the forcing case, the maximum amplitude is increased with a mean growth rate 
$2 \Delta$, see figure \ref{case4-rand} (right). 
Compared with the case  without initial random noise in figure \ref{fig-case4-theory1} 
when $\Delta=0.04$ the maximum amplitude grows but it requires a certain time interval 
for wave growth. Initial random noise stimulates the maximum growth rate to be earlier 
with the robust growth rate $2\Delta$.  
Instead of increasing $K$, we  fixed $K=4$ and increased  $\epsilon$.
We found that for $\epsilon=4$ the time $T_1 \approx 70$ 
for the amplitude to be stimulated but finite.
which is  much larger than $T_1$ for the case of smaller $\epsilon =1$.
 \\

\section{Discussion and summary}

In this paper we have used the forced NLS equation (\ref{fnls}) expressed in canonical form  
to model the generation of wave packets and breathers by adding a linear growth term
to the usual NLS equation.  In the absence of such forcing the principal solutions of the 
NLS equation are solitons and breathers, representing wave packets and 
{possibly rogue waves}, see \citet{KH09, OS10, gr13} for instance. 
In the forced NLS equation the forcing is 
represented  by a linear growth term with a rate parameter $\Delta$ so that 
$\Delta>0$ and $\Delta =0$ represents cases with and without forcing respectively,
see \citet{le07, tkpg08, mkmk13, br14, sl15, gr18} for the context of the 
generation of water waves by wind.  In this context the 
non-dimensional growth rate parameter $\Delta $ depends on several physical factors, 
especially the wind shear, the surface roughness and the initial water wave wavelength. 
It can range from $O(10^{-2})$ for weak winds to $O(10)$ for strong winds, see 
\citet{le07,tkpg08, sl15} for instance. 
Here we have varied $\Delta $ over the range 
from zero to order unity, covering the range of weak to moderate forcing
appropriate for our weakly nonlinear model. 

Four scenarios are investigated through an appropriate choice of initial condition.
(1) an initial condition which in the unforced case would generate a Peregrine breather (\ref{PB}); 
 (2) an initial condition which in the unforced case would generate a moving soliton (\ref{SOL}); 
 (3) a slowly-varying long wave perturbation which in the unforced case would generate either a few solitons for $\epsilon$ of order unity, or a family of Peregrine breathers 
 when $\epsilon $ becomes very small; 
 (4) a long-wave periodic perturbation which in the unforced case would generate 
 modulation instability and the formation of  both solitons and breathers.  
 
 In case (1) a Peregrine breather is formed  when $\Delta =0$ and agrees with 
  the well-known exact solution. When $\Delta >0$ a forced Peregrine breather  initially develops with 
  an increased amplitude growing at twice the linear growth rate, but instead of decreasing to zero, 
  the amplitude continues to grow and oscillates with increasing frequency.
  In case (2) with $\Delta=0$ a steadily moving soliton with a constant 
  amplitude forms. When $\Delta>0$ the soliton amplitude grows at the rate $2\Delta $, twice the linear growth rate
  while continuing to move with a constant speed.  In case (3) with $\Delta=0$ and with a 
  very small dispersion parameter  $\epsilon=1/33$, there is a gradient catastrophe 
  followed by the formation of a family of Peregrine breathers as expected, see \citet{gr13}.
  When $\Delta >0$, in contrast to the unforced case the Peregrine breathers are replaced 
  by a mixture of breathers and solitons.
  Three scenarios were found, the generation of mainly moving solitons with increasing amplitudes, 
  nearly stationary   solitons with increasing amplitude, and a combination of both 
  breathers and increasing amplitude solitons. 
 In case (4) a periodic long-wave perturbation with wavenumber $K$ is imposed as the initial condition.
 Modulation instability with wavenumber $K$ occurs when $\epsilon K < \sqrt{2} |M| $ 
 where $M$ is the  initial amplitude of the periodic long wave. 
 When $\Delta =0$ a mixture of solitons and breathers form as is well-known, 
 see \citet{OS10} for insatnce.   However, as the forcing parameter $\Delta $ increases,  
 the breathers begin to be eliminated and are replaced by solitons with growing amplitudes, 
  progressively fewer  forming as the  forcing increases. 

For each case (1-4) of these initial conditions, we investigated the effect of  initial random noise. 
The case of an initial moving soliton is unchanged except that some small solitons are generated 
downstream ($X>0$) due to the initial random noise perturbation. 
The maximum growth rate $2 \Delta$ can still be used to 
make an accurate prediction of the amplitude growth.
For the cases when  breathers form, initial random noise shifts the locations of the 
unforced solutions. When forcing is involved, the maximum growth rate of the breathers 
 increases and oscillates with a mean growth rate $2\Delta$. 
 For the case of initial periodic plane wave, initial random noise changes the modulation pattern from 
 deterministic to chaotic with  the implication that the location of the maximum amplitude cannot 
 be determined exactly. However, the growth of the maximum amplitude is still approximately $2\Delta$. 
 Overall, the predicted growth rate $2\Delta$ is robust for these initial value problems 
 with and without an initial random noise effect.  
  
Modulation instability and the subsequent formation of small amplitude waves that 
generate large amplitude wave or sometimes rogue wave has been studied 
experimentally for water waves  by many authors, see for instance \cite{Ono05}. 
Large-amplitude waves are subject to modulation instability,
measured  by the Benjamin-Feir index $\hbox{BFI}$, 
 the ratio of wave steepness (nonlinearity) to spectral bandwidth (dispersion).
Here  $\hbox{BFI} \approx 1/\epsilon$, see \citet{gr13}.
Even for $\hbox{BFI}=1$ which is a moderate value, forcing stimulates modulation instability. 
When the forcing term $\Delta>0$, there are two stages: the first stage is the development of  
breathers which could be interpreted as a random sea state as  time evolves,  and then a  
second stage forms with  large amplitude waves. The large waves in the second stage 
collect energy from neighbouring small waves with different wave frequencies. 
Rogue waves are observed  when $\hbox{BFI}$ is large with amplitudes three or four times the
 background sea state during their evolution, see for instance \cite{Ono05, OS10}. 
 In our present work even for $\hbox{BFI}=1$ modulation instability occurs in the predicted 
 long-wave perturbation range ($\epsilon K < 2 |M|$) in the first stage, but then large amplitude waves 
 develop due to the forcing. The larger the value of the forcing parameter, 
 the larger are the  waves in the second stage and they  become unstable.
Instead of using the periodic wave plane as an initial condition, 
in case (3)  the initial condition of a slowly-varying long wave perturbation 
with a $\hbox{sech}$-profile also  develops into  modulation instability. 

Recent work by \cite{Le20} shows the region of high and low wave frequency 
nonlinear wave interaction where a nonlinear wave component can grow exponentially, 
leading to rogue wave packets. Outside this region, the small waves are stable. 
This situation is comparable with our results shown in figure \ref{fig-sech-gam05-ep02}  
with and without forcing effects,  interpreted here as a wind effect.
It can be seen from the case  without forcing  that a sequence of breathers is generated 
as expected but the introduction of  forcing  can generate large waves 
growing in wave amplitude and stationary. An explicit formula to express for 
 rogue wave formation under forcing  and nonlinear wave packet interaction remains challenged for further studies. \\ \\


\bigskip
\noindent
{\bf Acknowledgments}: 
RG was supported by the Leverhulme Trust through the award of a Leverhulme Emeritus Fellowship EM-2019-030.

\bibliographystyle{apalike}

\bibliography{fNLS}


\begin{figure}[ht]
\begin{center}
\includegraphics[width=9cm, height=7cm]{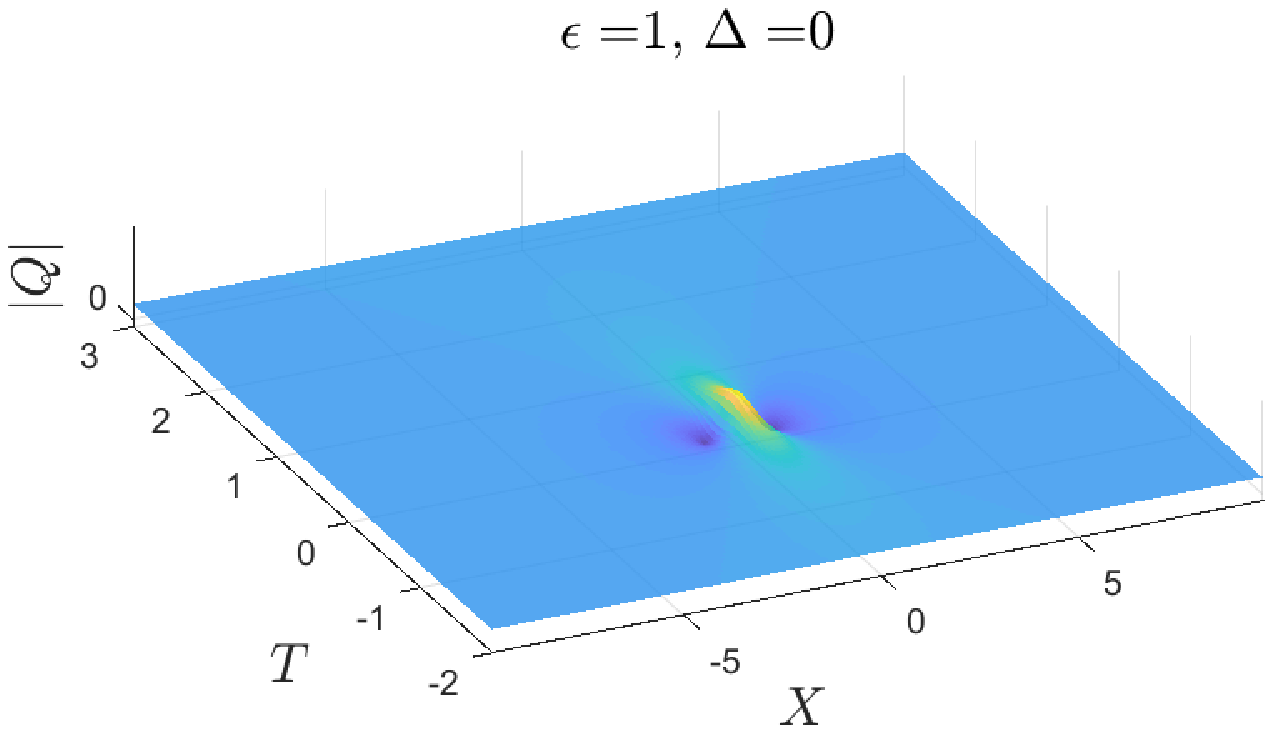}
\caption{Case 1:  Surface plot of the unforced Peregrine breather (\ref{PB}) when $\epsilon=1.0$, $M=1$, $\Delta=0.0$ }
\label{PB1}
\end{center}
\end{figure}

\begin{figure}[ht]
\begin{center}
\includegraphics[width=9cm, height=7cm]{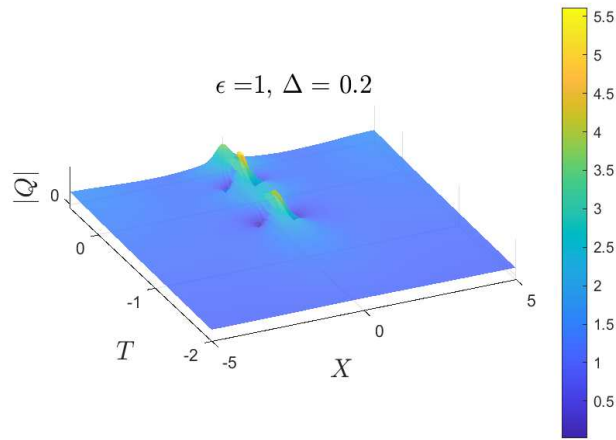}
\caption{Case 1:  Surface plot of the forced Peregrine breather when $\epsilon=1.0$, $M=1$, $\Delta=0.2$ 
with the initial condition (\ref{PB}) at $T_0 = -2$.}
\label{PB3}
\end{center}
\end{figure}

\begin{figure}[ht]
\begin{center}
\includegraphics[width=9cm, height=6.5cm]{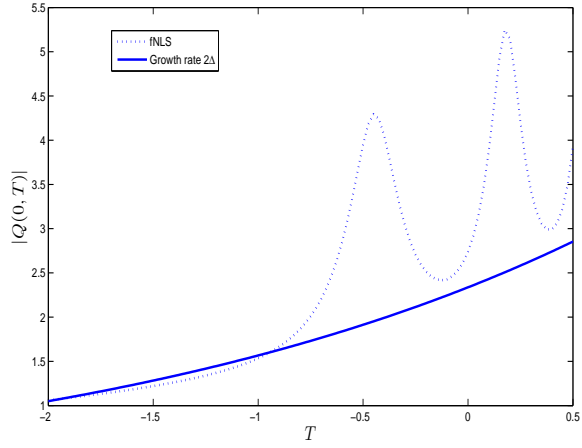}
\caption{Case 1: $|Q(0,T)|$ when $\epsilon=1.0$, $M=1$, $\Delta=0.2$ 
with the initial condition that (\ref{PB}) at $T_0 = -2$.}
\label{PB5}
\end{center}
\end{figure}

\begin{figure}[ht]
\begin{center}
\includegraphics[width=9cm, height=7cm]{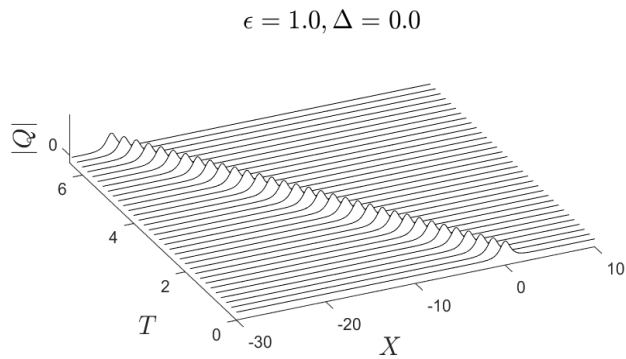}
\caption{Case 2: Constant amplitude moving soliton from equation (\ref{fnls}) when $\epsilon=1.0, M=2, K=-2$, and $\Delta=0.0$ }
\label{fig2-1}
\end{center}
\end{figure}

\begin{figure}[ht]
\begin{center}
\includegraphics[width=9cm, height=7cm]{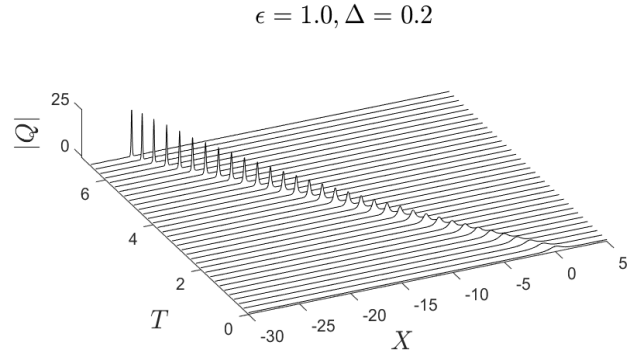}
\caption{Case 2: Growing amplitude moving soliton from equation (\ref{fnls}) when $\epsilon=1.0, M=2, K=-2$, and $\Delta=0.2$ }
\label{fig2-2}
\end{center}
\end{figure}

\begin{figure}[ht]
\begin{center}
\includegraphics[width=9cm, height=7cm]{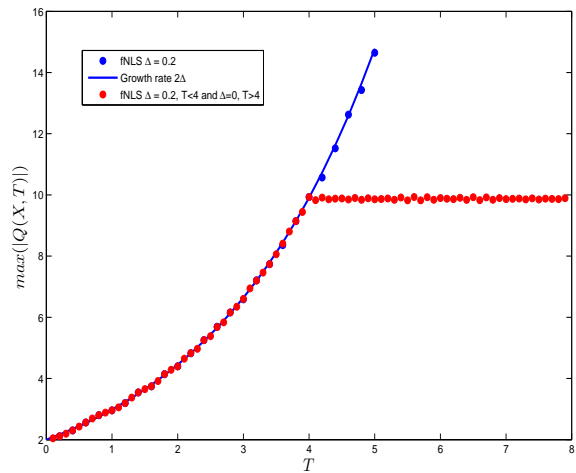}
\caption{Case 2: Maximum of $|Q(X,T)|$ and growth rate $2\Delta$ when $\epsilon=1.0, M=2, K=-2$, and $\Delta=0.2$, forcing is turned off when $T>4$. }
\label{fig-case2-theory}
\end{center}
\end{figure}

\begin{figure}[ht]
\begin{center}
\includegraphics[width=9cm, height=7cm]{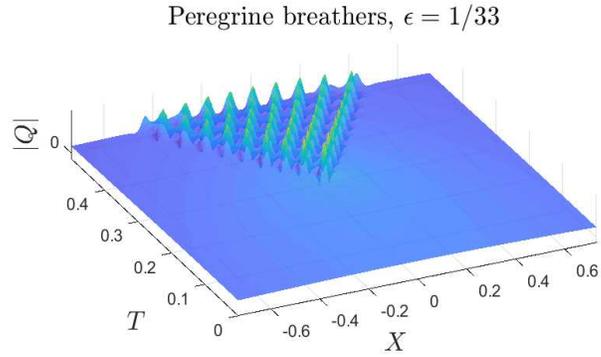}
\caption{Case 3: A family of Peregrine breathers generated from the initial condition (\ref{PBsic}) when 
$\epsilon=1/33, \gamma=1, M=1$  and  $\Delta=0$. }
\label{PB-surf}
\end{center}
\end{figure}  

\begin{figure}[ht]
\begin{center}
\includegraphics[width=9cm, height=7cm]{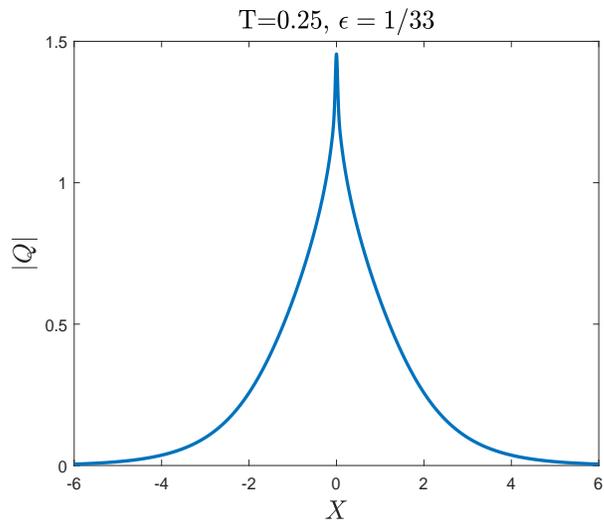}
\caption{Case 3:  The gradient catastrophe generated from the initial condition (\ref{PBsic})  
when $\epsilon=1/33, \gamma=1, M=1$  and $\Delta=0$ }
\label{PB-gradient}
\end{center}
\end{figure}  

\newpage
\begin{figure}[ht] 
\begin{center}
\begin{tabular}{cc}
\includegraphics[height=5cm,width=8cm]{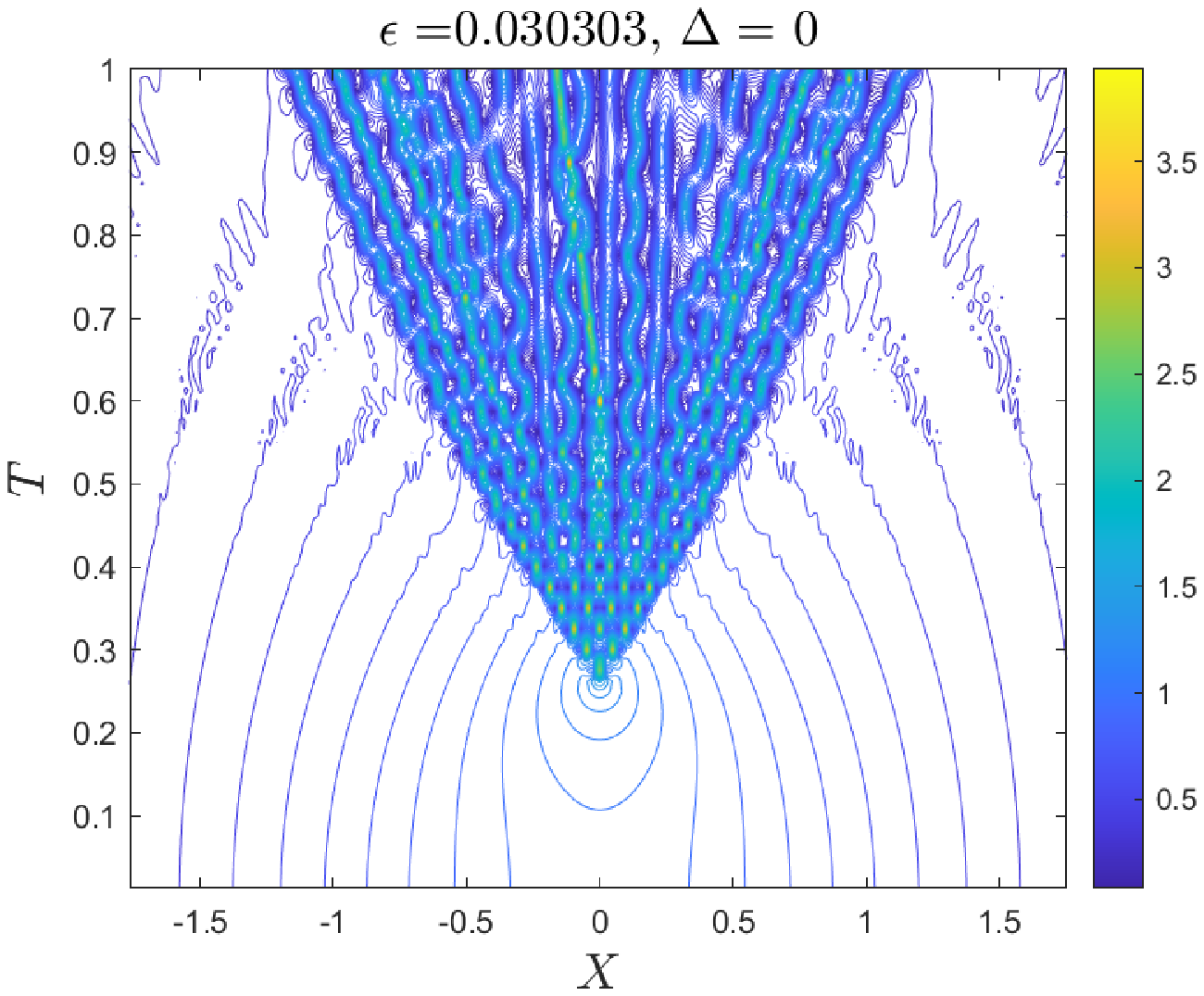} 
\includegraphics[height=5cm,width=8cm]{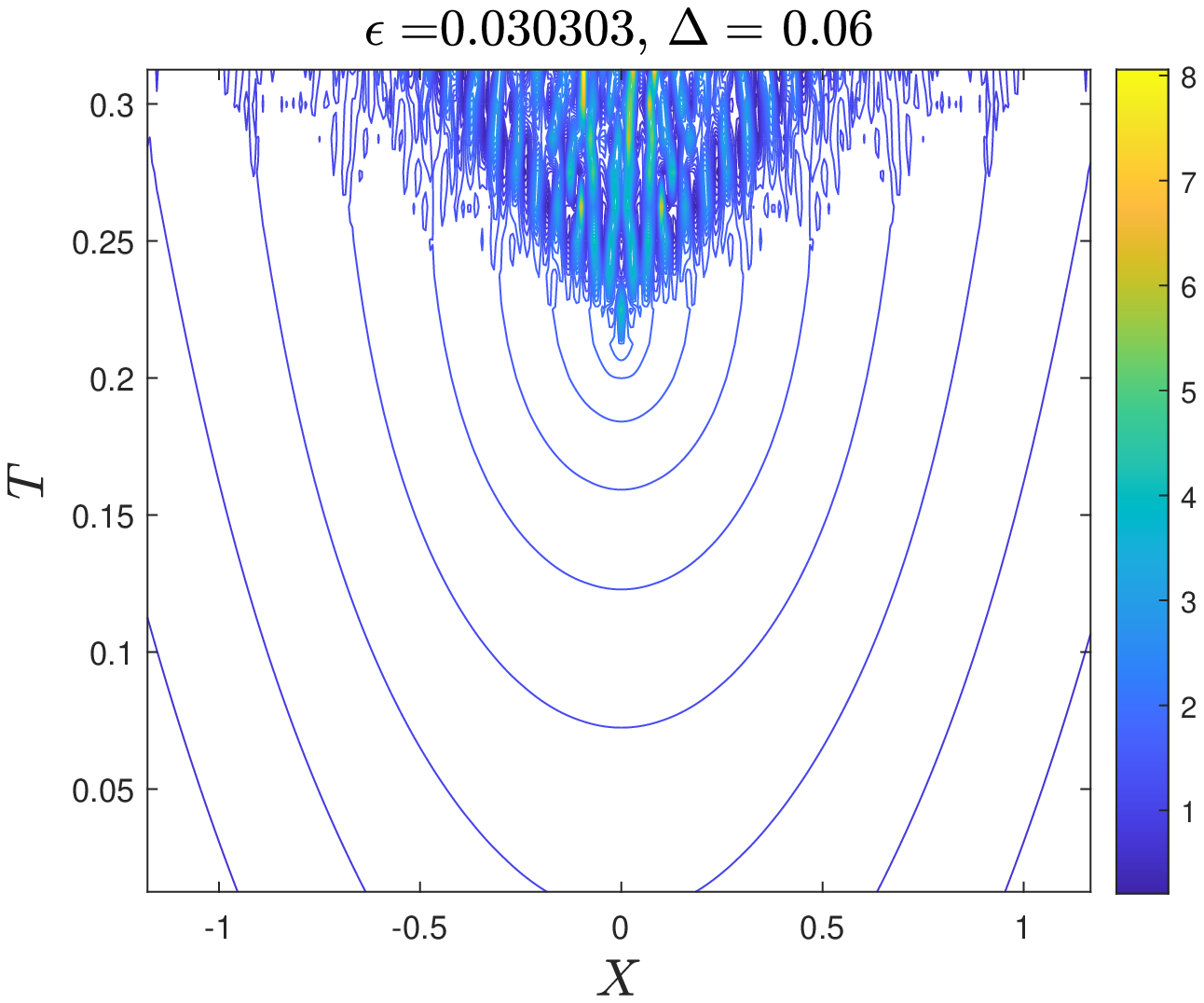} \\
\end{tabular}
\caption{Case 3: The initial condition is   (\ref{PBsic}) with 
$\gamma = 1.0$, $\epsilon=1/33, M=1$  and $\Delta = 0$, $\Delta = 0.06$.}
\label{fig-sech-gam1-ep003-a}
\end{center}
\end{figure}

\newpage

\begin{figure}[ht]  
\begin{center}
\begin{tabular}{cc}
\includegraphics[height=5cm,width=8cm]{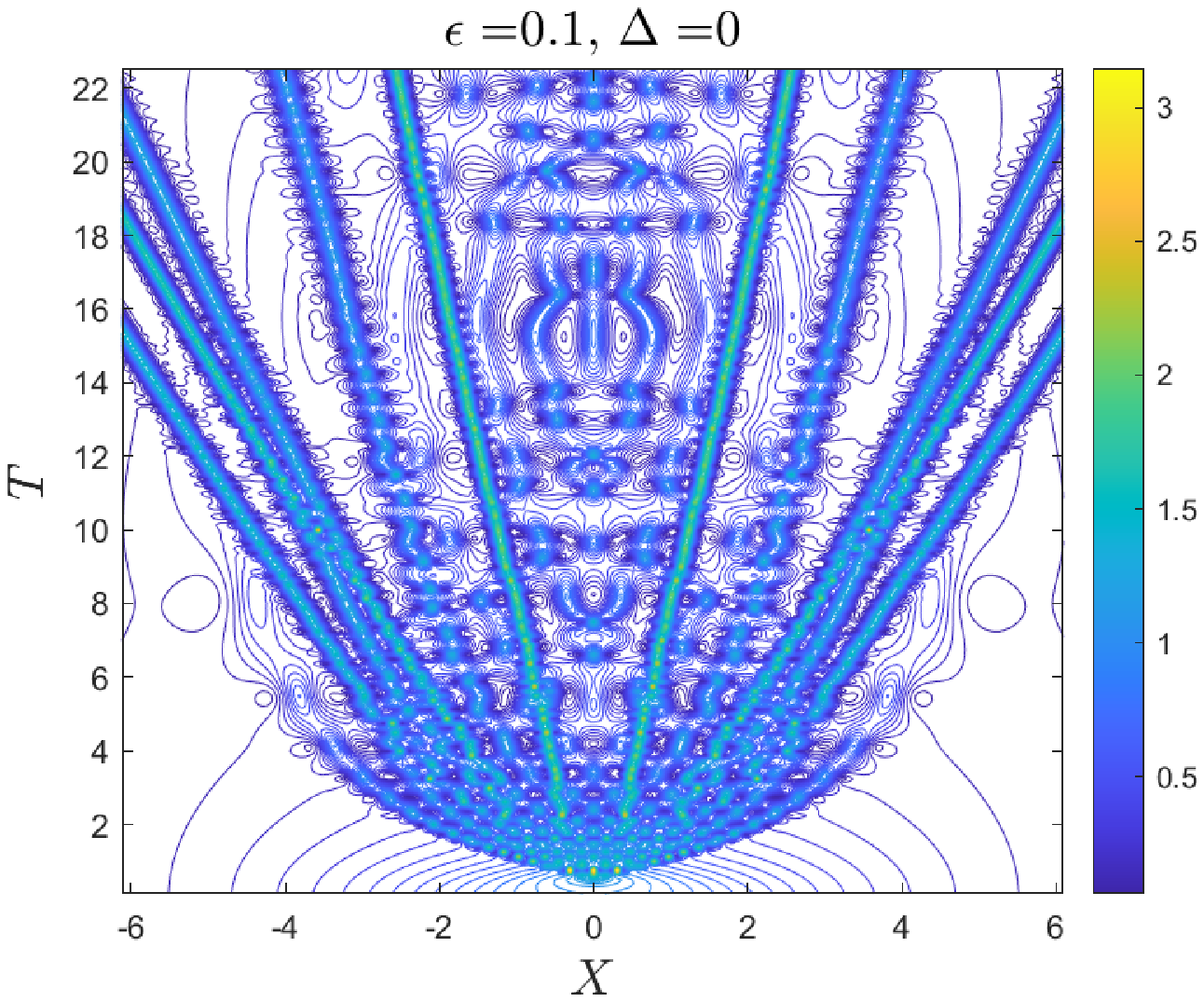} 
\includegraphics[height=5cm,width=8cm]{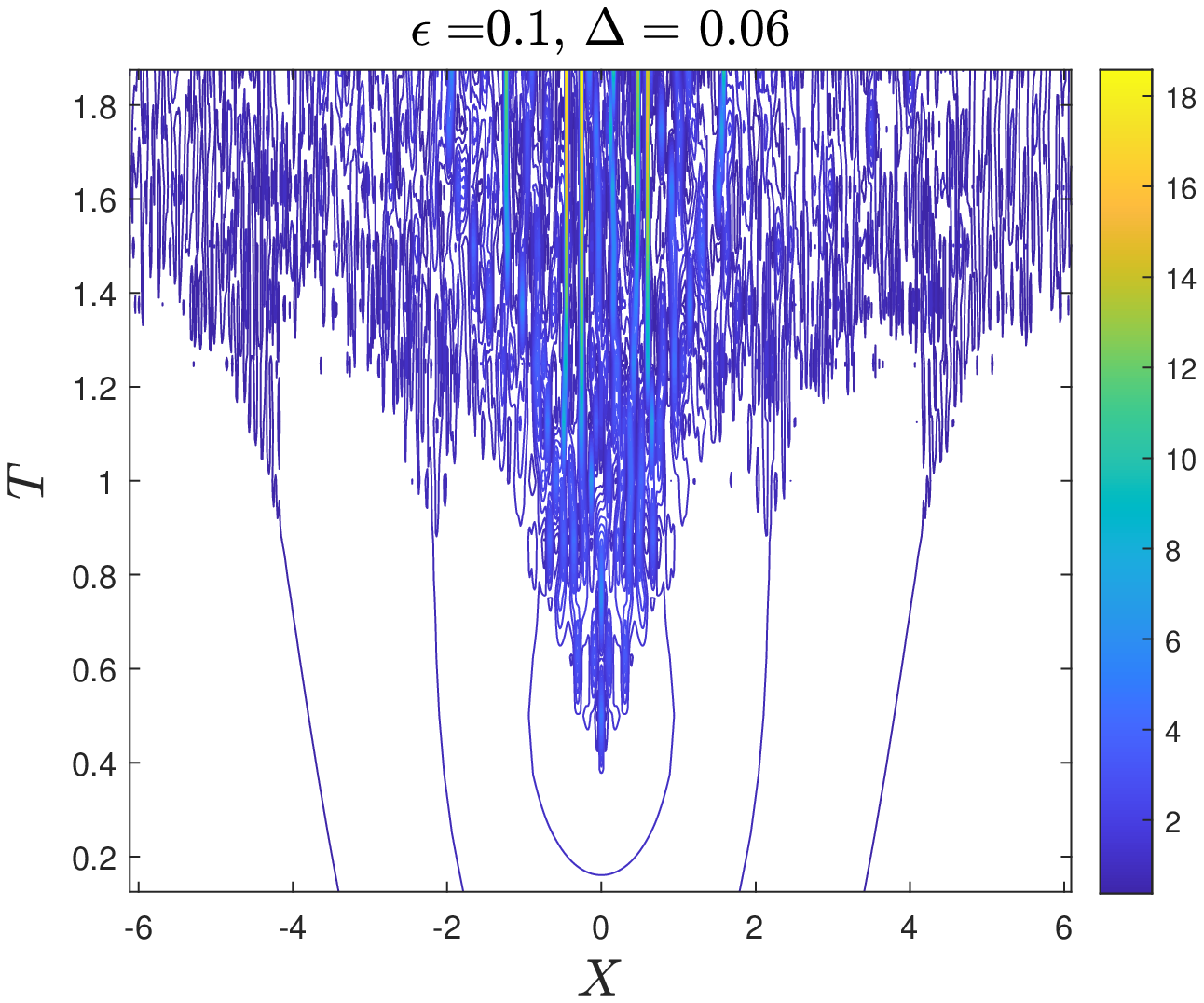} \\
\end{tabular}
\caption{Case 3: The initial condition is  (\ref{PBsic}) with 
$\gamma = 0.5$, $\epsilon=0.1, M=1$  and $\Delta = 0$, $\Delta = 0.06$.}
\label{fig-sech-gam05-ep01}
\end{center}
\end{figure}

\newpage
\begin{figure}[ht] 
\begin{center}
\begin{tabular}{cc}
\includegraphics[height=5cm,width=8cm]{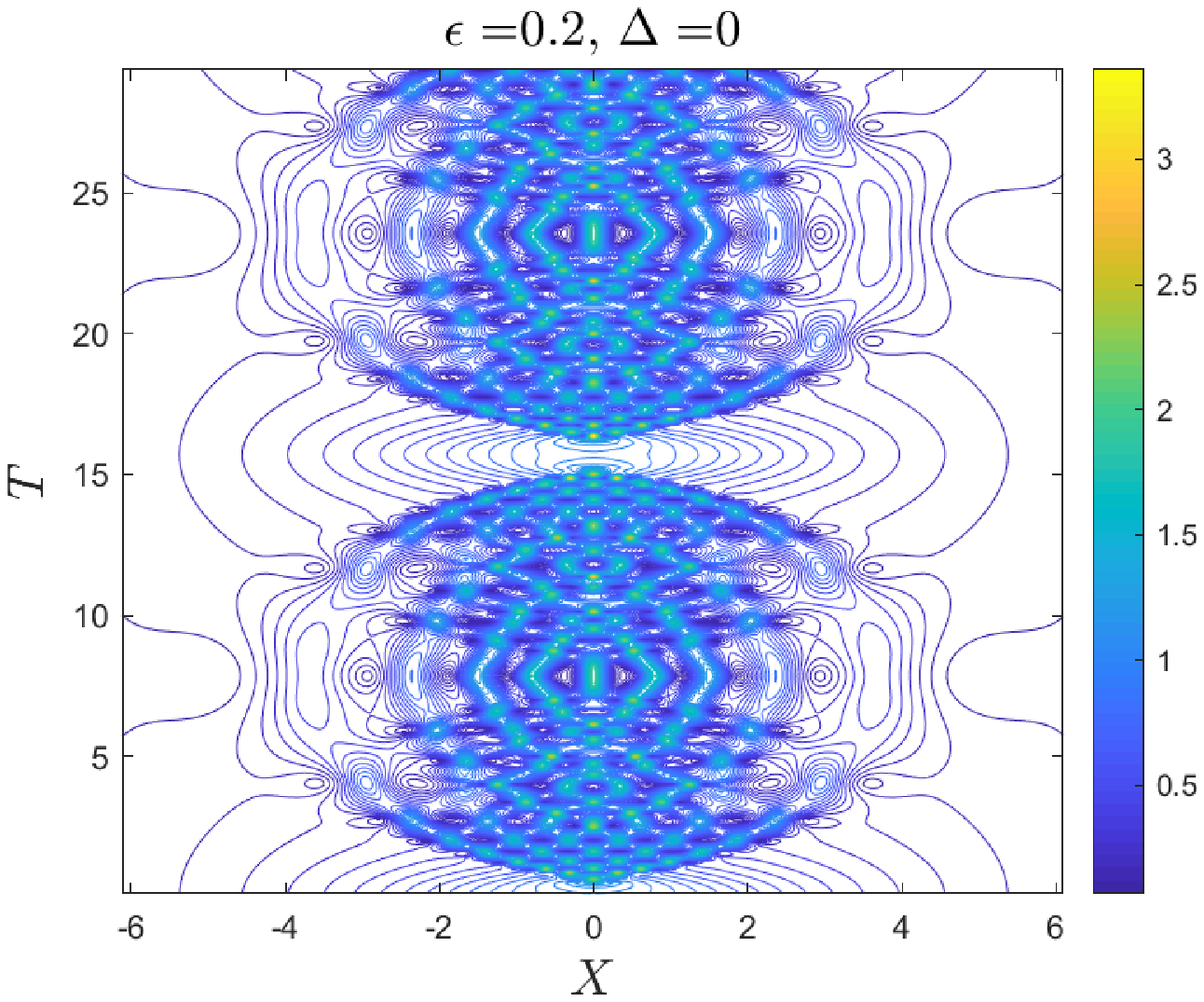} 
\includegraphics[height=5cm,width=8cm]{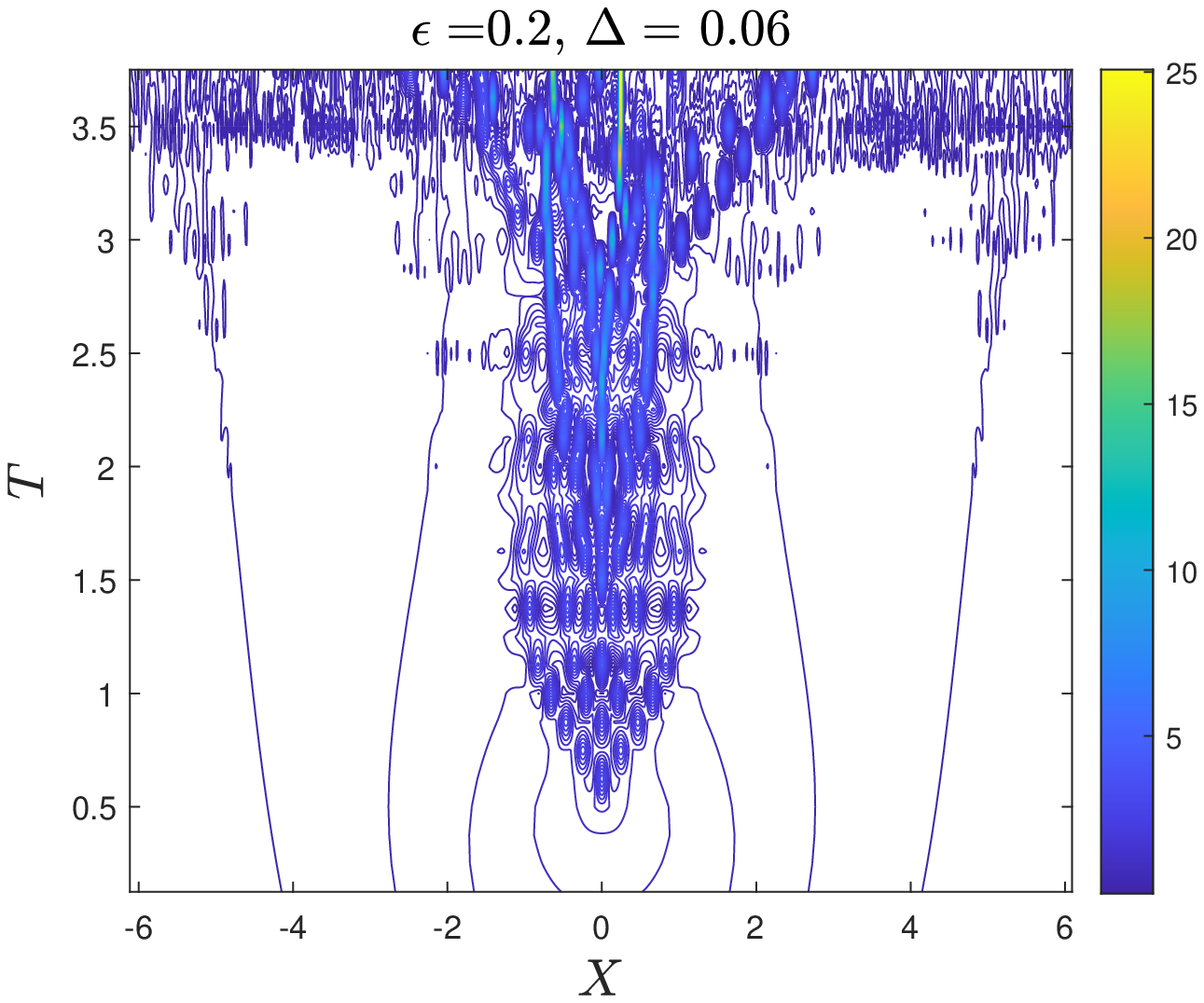} \\
\end{tabular}
\caption{Case 3: The initial condition is  (\ref{PBsic}) with 
$\gamma = 0.5, \epsilon=0.2, M=1$  and $\Delta = 0$, $\Delta = 0.06$.}
\label{fig-sech-gam05-ep02}
\end{center}
\end{figure}

\newpage
\begin{figure}[ht] 
\begin{center}
\begin{tabular}{cc}
\includegraphics[height=5cm,width=8cm]{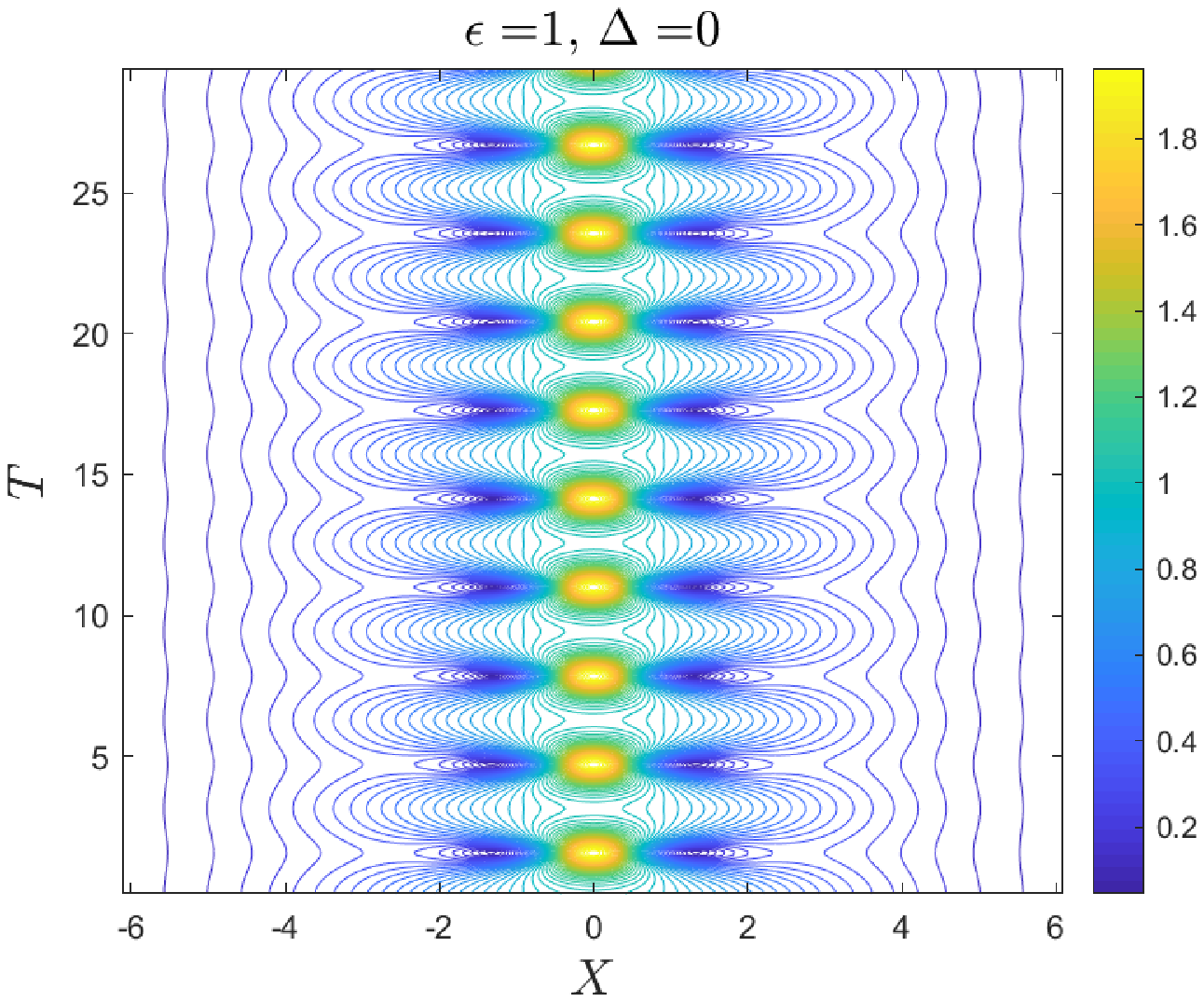} 
\includegraphics[height=5cm,width=8cm]{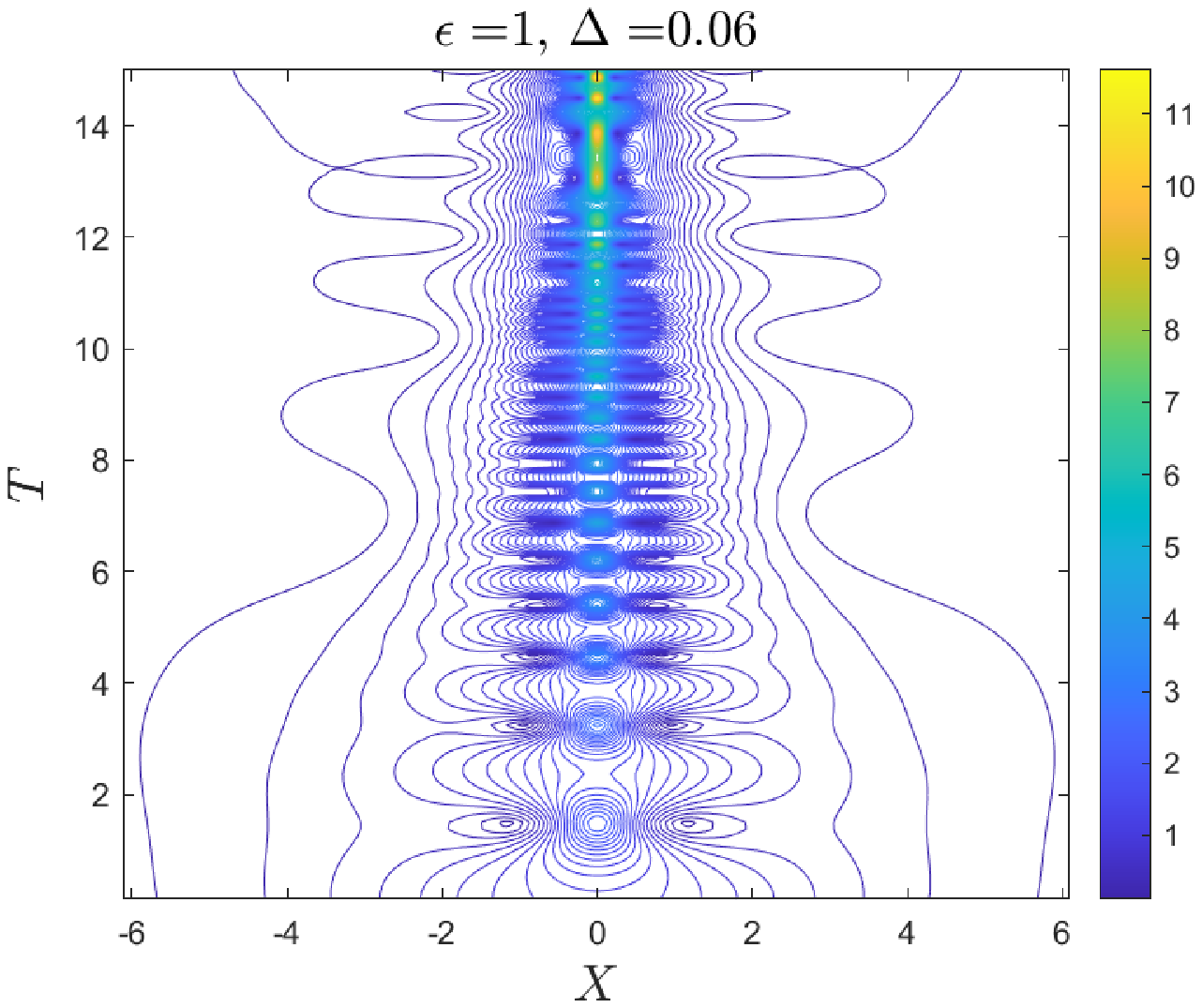} \\
\end{tabular}
\caption{Case 3: The initial condition is  (\ref{PBsic}) with 
$\gamma = 0.5,\epsilon=1.0, M=1$  and $\Delta = 0$, $\Delta = 0.06$.}
\label{fig-sech-gam05-ep1}
\end{center}
\end{figure}

\begin{figure}[ht]
\begin{center}
\includegraphics[width=12cm, height=8cm]{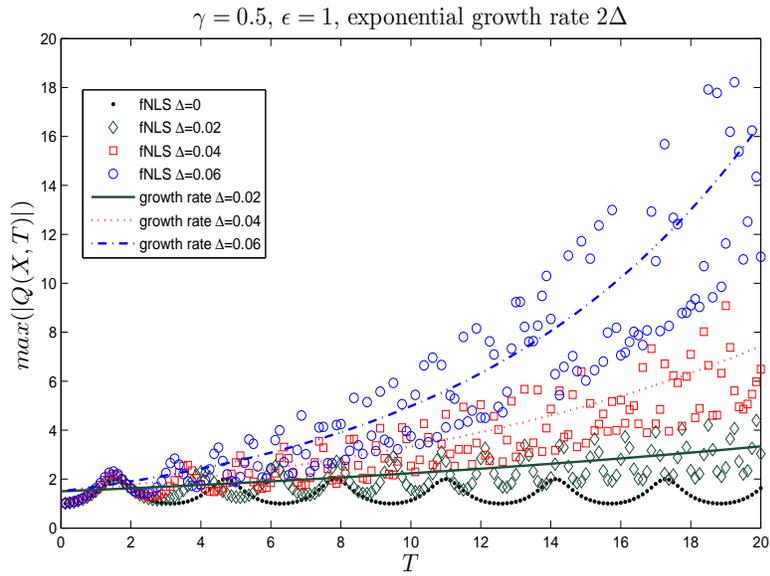}
\caption{Case 3: Maximum of $|Q(X,T)|$ and growth rate $2\Delta$ when 
$\gamma = 0.5, \epsilon=1.0, M=1$ for various values of $\Delta$.}
\label{fig-case3-theory}
\end{center}
\end{figure}

\begin{figure}[ht]
\begin{center}
\includegraphics[width=12cm, height=8cm]{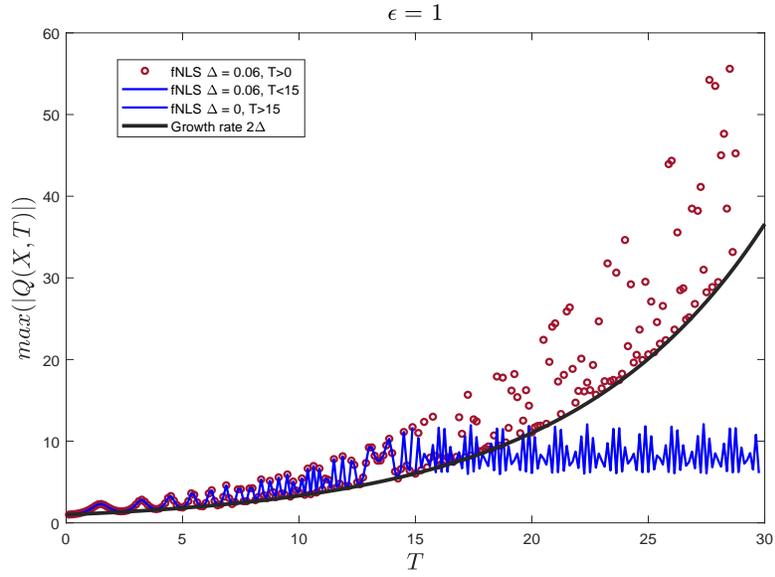}
\caption{Case 3: Maximum of $|Q(X,T)|$ and growth rate $2\Delta$ when 
$\gamma = 0.5, \epsilon=1.0, M=1$ and $\Delta=0.06$. 
The forcing is turned off when $T>15$.}
\label{fig-case3-theory-turn}.
\end{center}
\end{figure}

\newpage
\begin{figure}[ht]
\begin{center}
\begin{tabular}{cc}
\includegraphics[height=5cm,width=8cm]{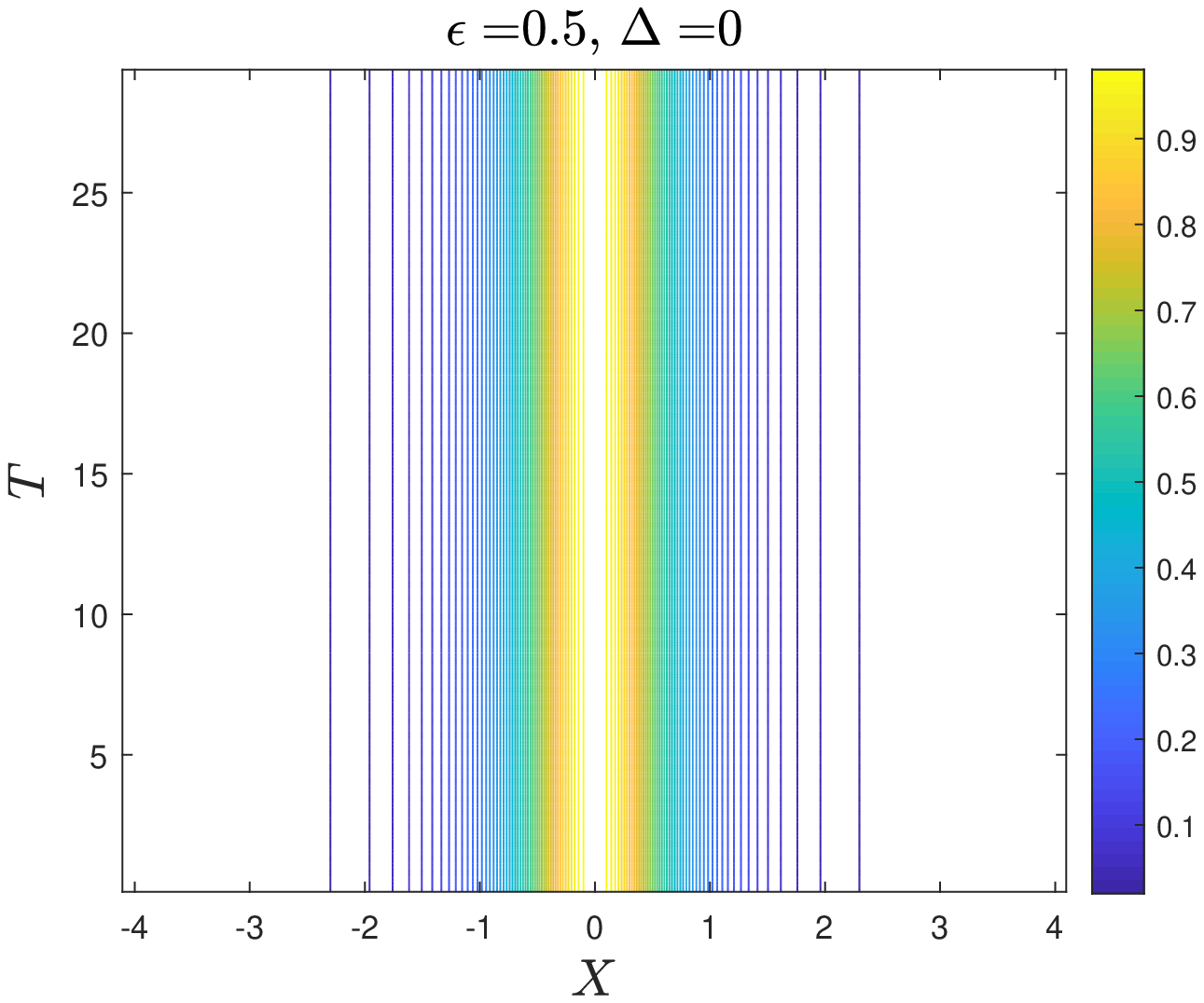} 
\includegraphics[height=5cm,width=8cm]{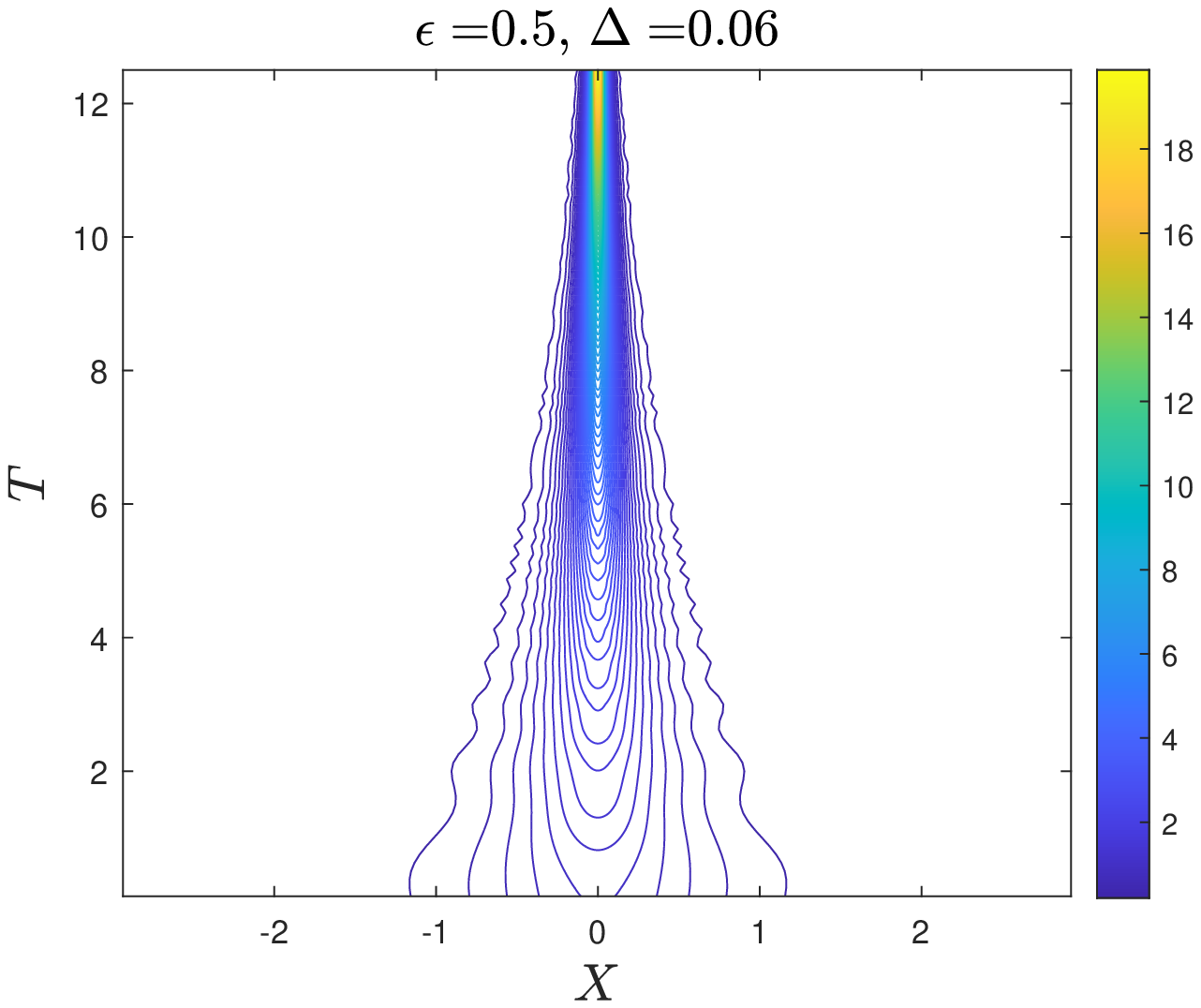} \\
\end{tabular}
\caption{Case 3: The initial condition is   (\ref{PBsic}) with 
$\gamma = 2.0$, $\epsilon=0.5, M=1$  and $\Delta = 0$, $\Delta = 0.06$.}
\label{fig-sech-gam2-ep05}
\end{center}
\end{figure}

\newpage
\begin{figure}[ht]  
\begin{center}
\begin{tabular}{cc}
\includegraphics[height=5cm,width=8cm]{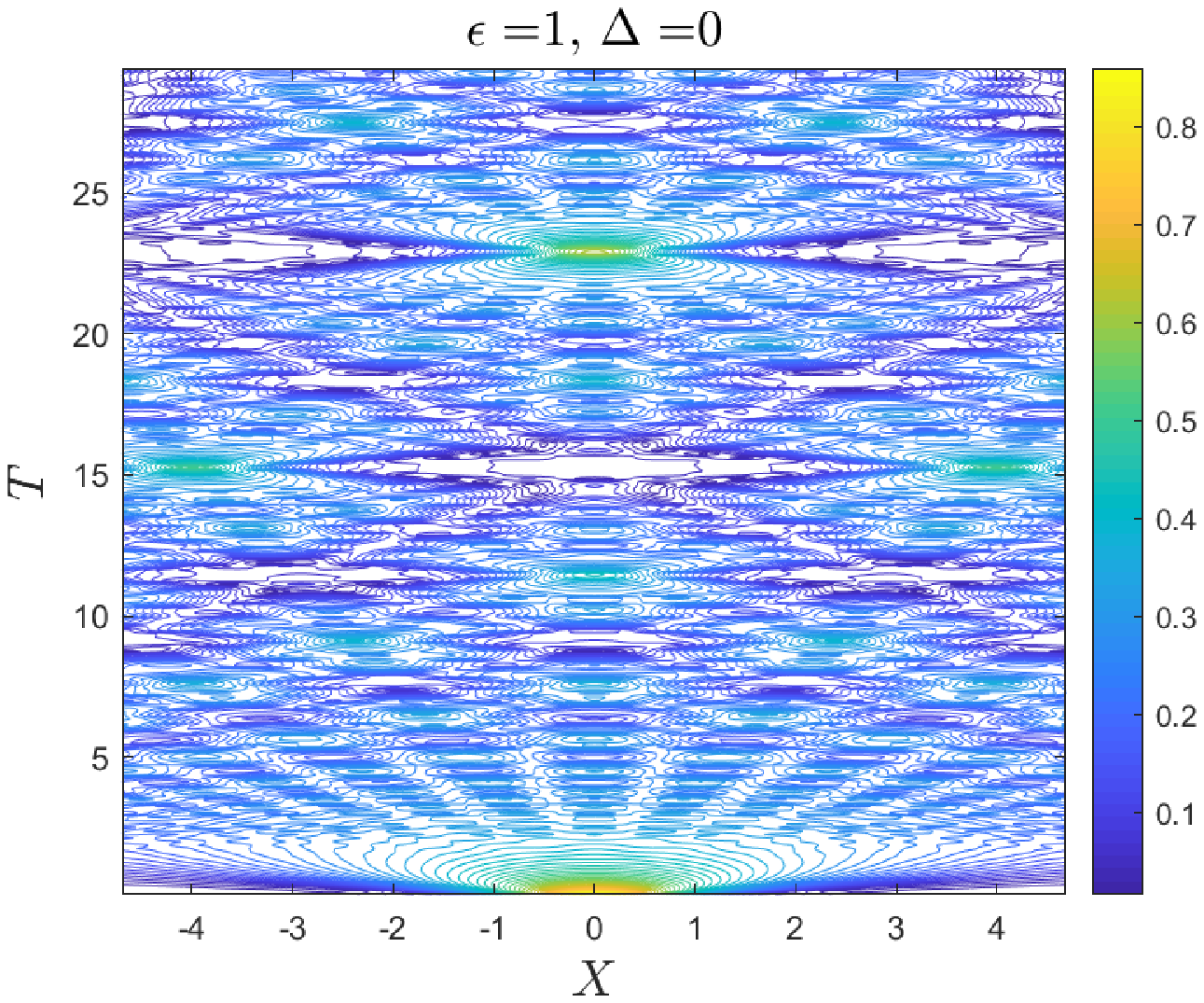} 
\includegraphics[height=5cm,width=8cm]{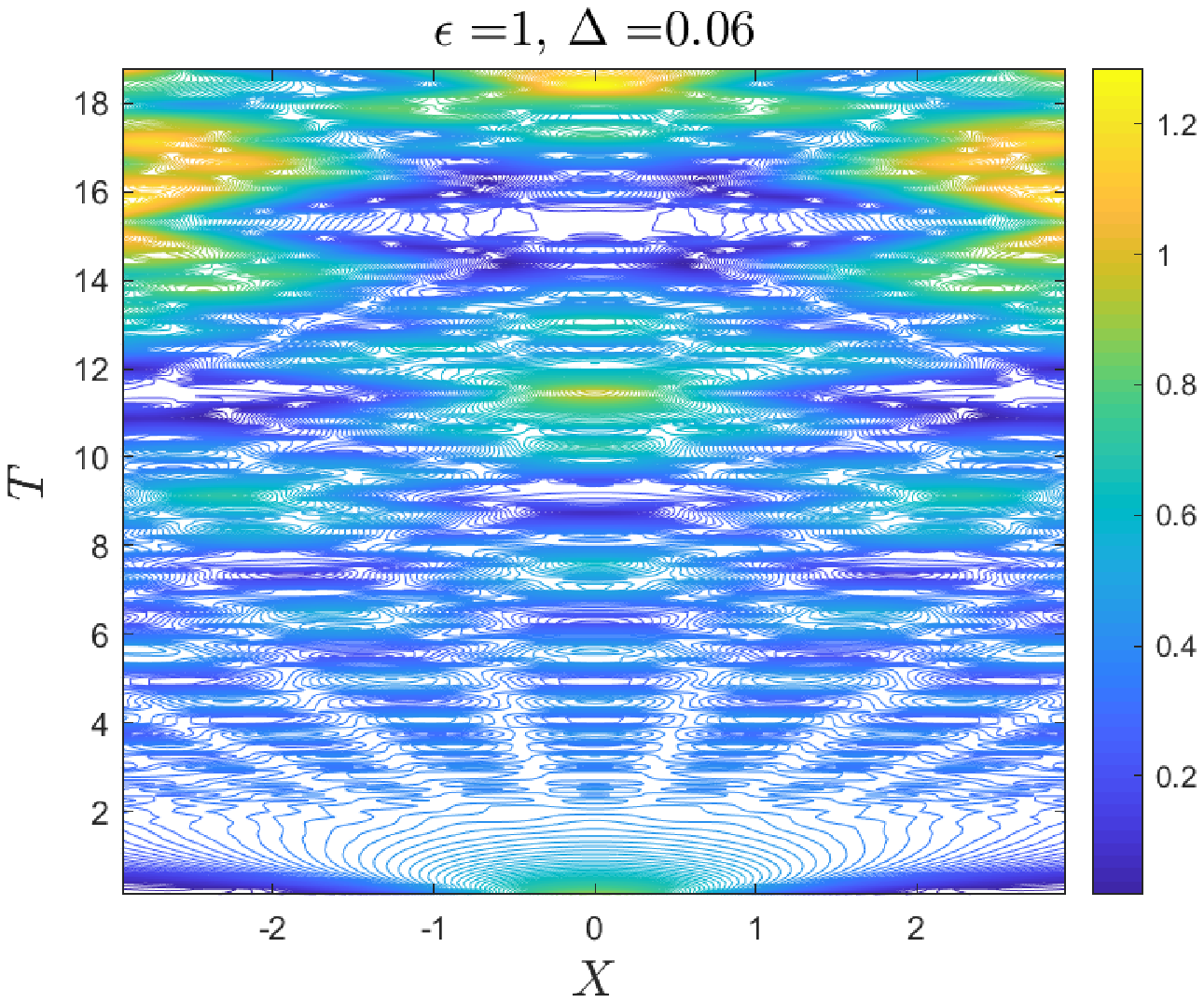} \\
\end{tabular}
\caption{Case 3: The initial condition is (\ref{PBsic}) with 
$\gamma = 2.0, \epsilon=1.0, M=1$  and $\Delta = 0$, $\Delta = 0.06$.}
\label{fig-sech-gam2-ep1}
\end{center}
\end{figure}

\begin{figure}[ht]  
\begin{center}
\includegraphics[height=5cm,width=8cm]{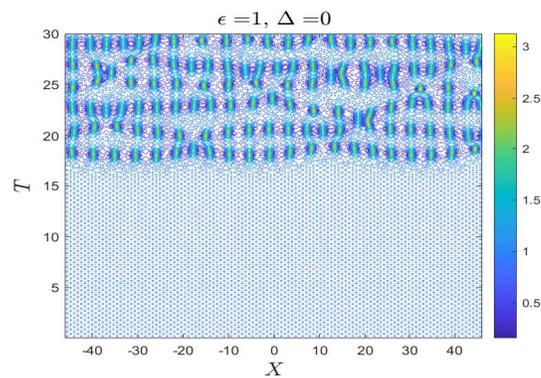} 
\caption{Case 4: The initial condition is (\ref{MIic}) with 
$\epsilon=1.0, M=1, K=4$  and $\Delta = 0$.}
\label{fig-ep4a}
\end{center}
\end{figure}

\begin{figure}[ht] 
\begin{center}
\begin{tabular}{cc}
\includegraphics[height=5cm,width=8cm]{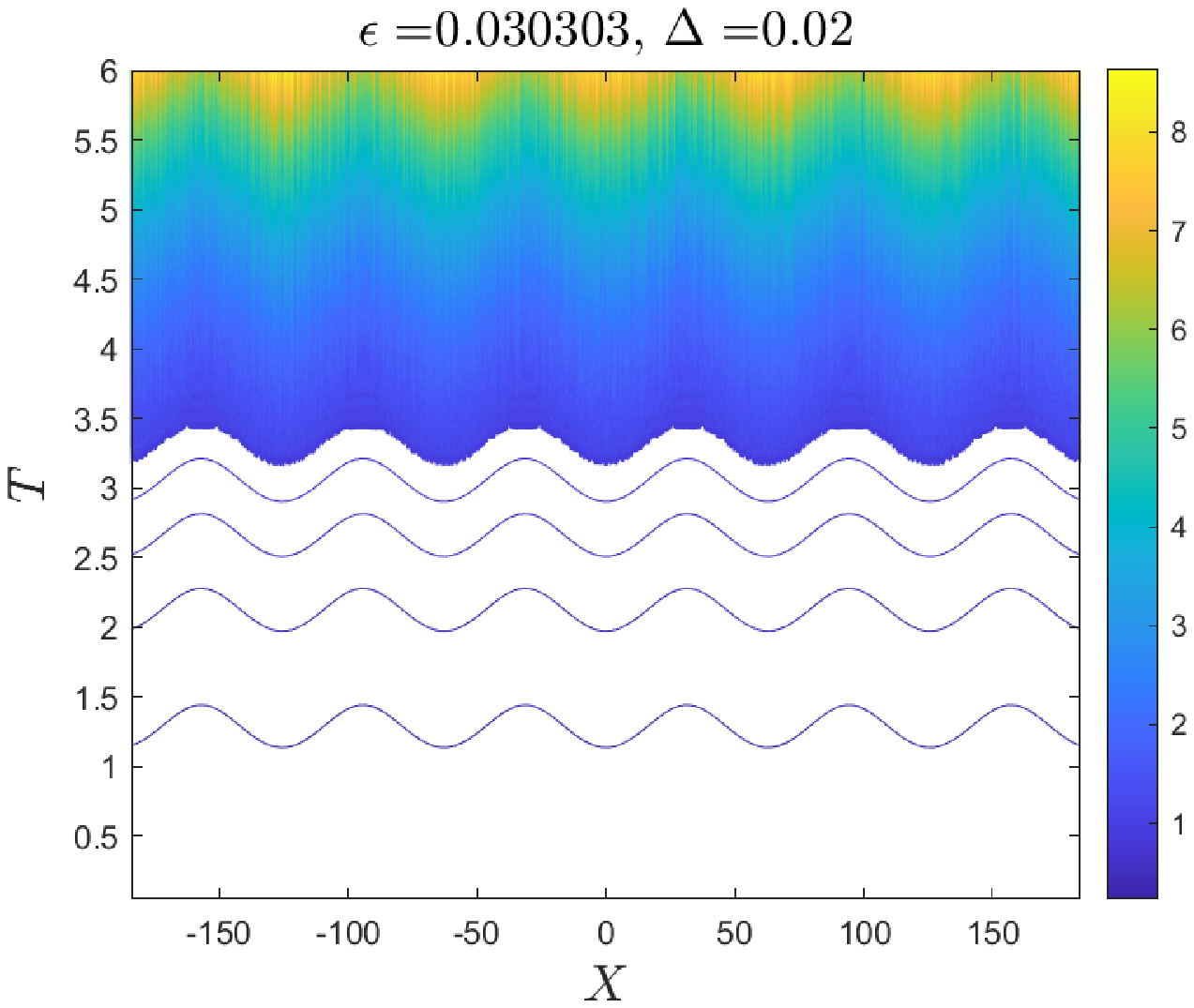} &
\includegraphics[height=5cm,width=8cm]{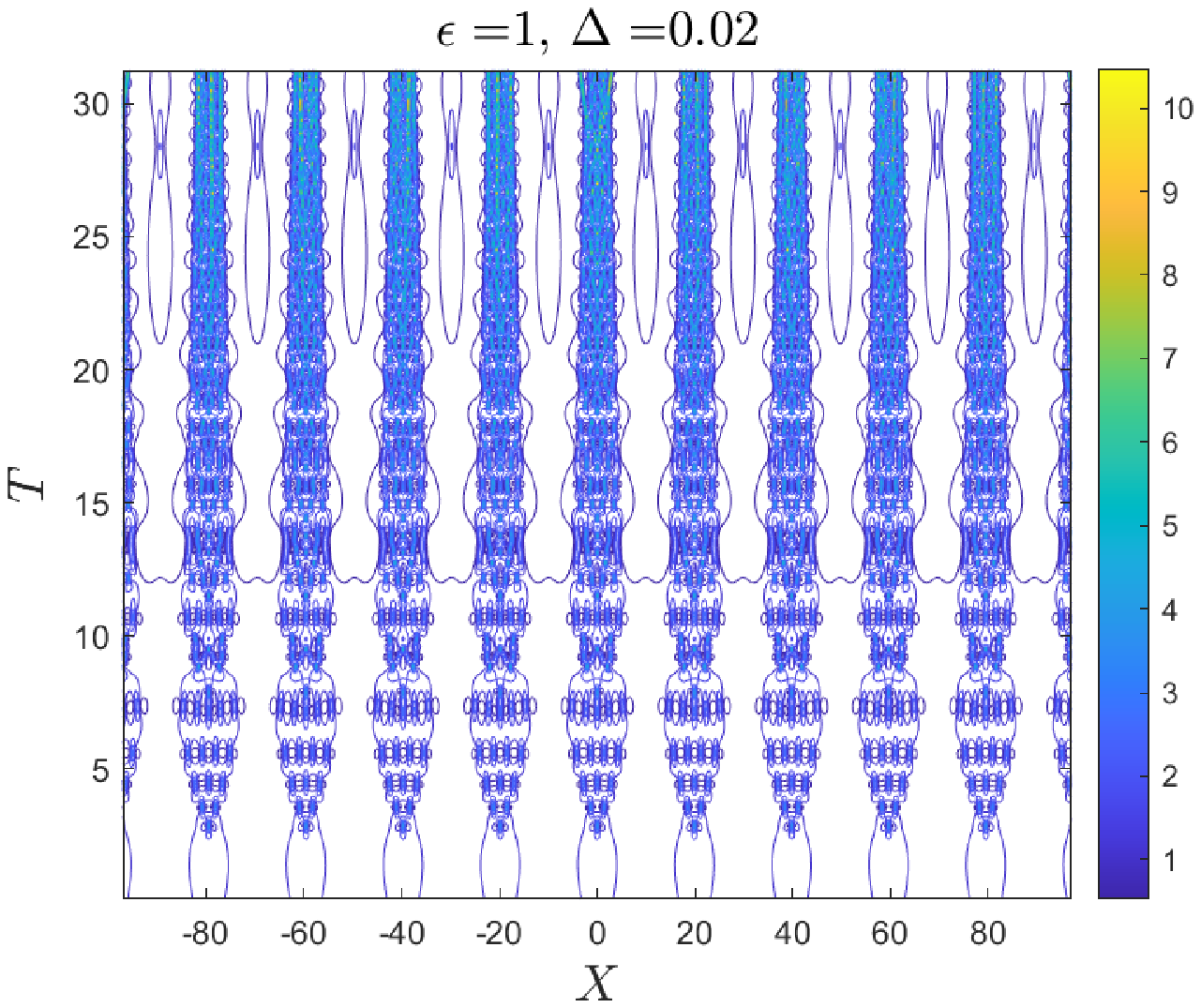} \\
\includegraphics[height=5cm,width=8cm]{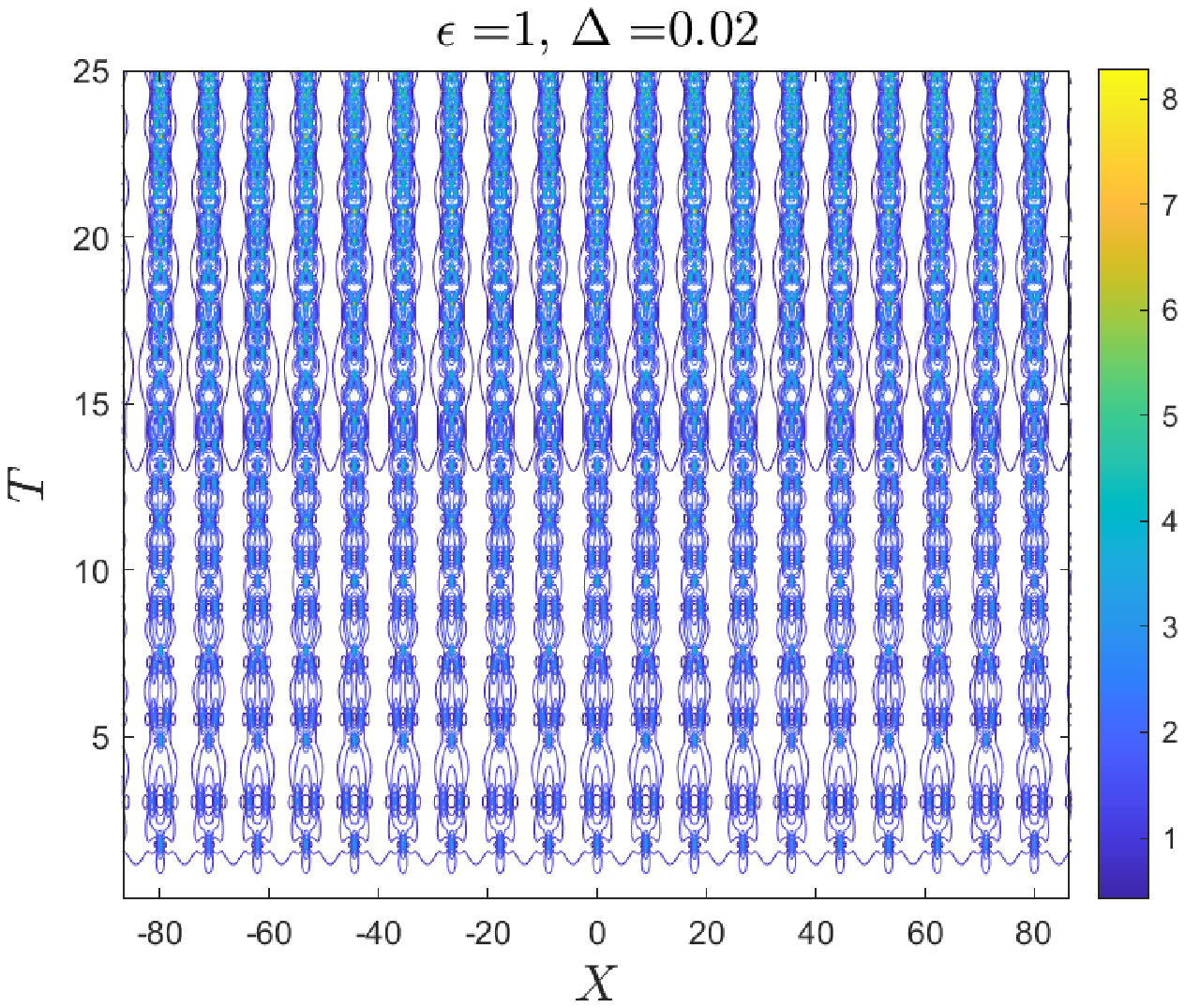} &
\includegraphics[height=5cm,width=8cm]{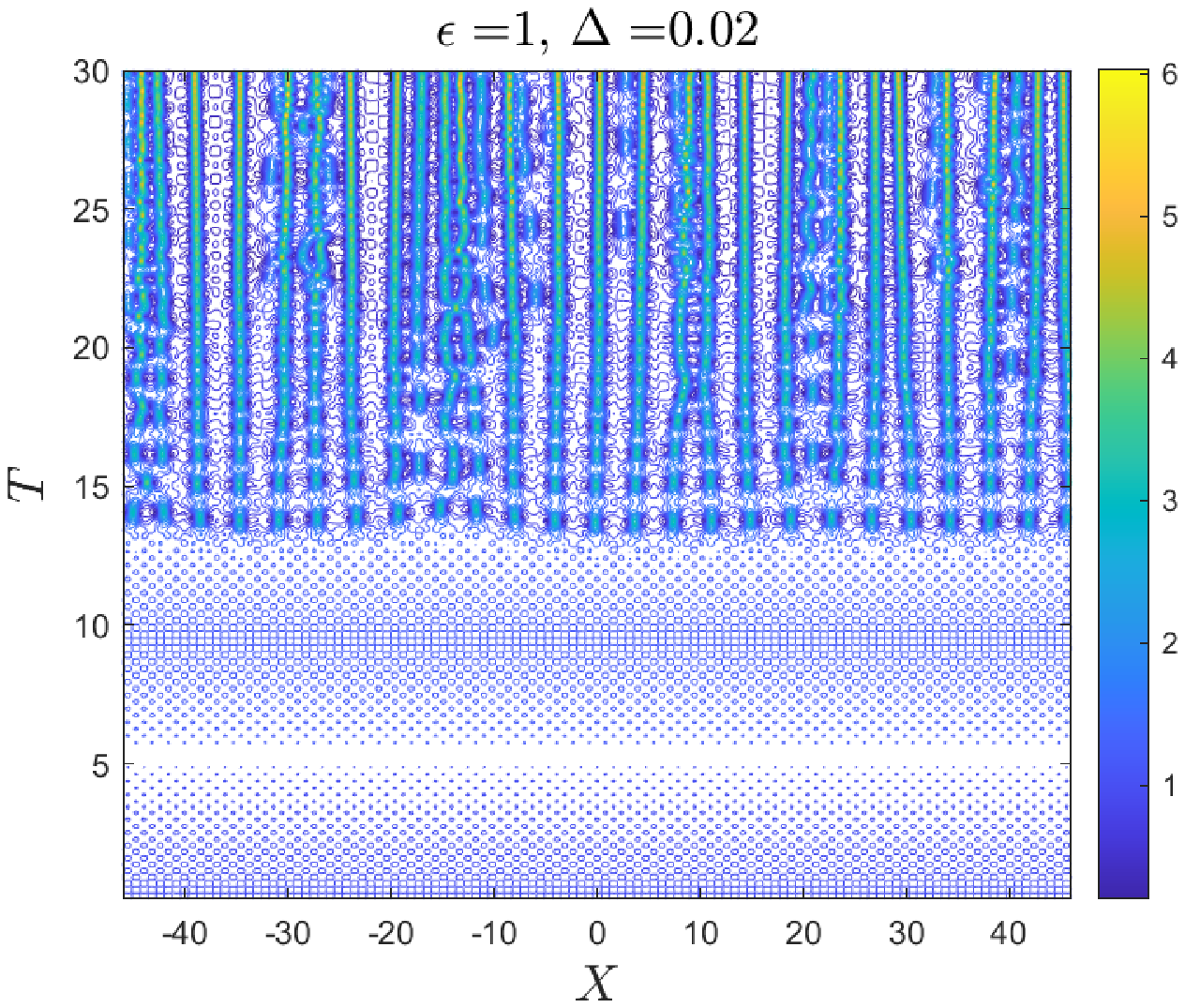} \\
\end{tabular}
\caption{Case 4: The initial condition is (\ref{MIic}) with  $\Delta=0.02$.  
Contour plots of $|Q|$ when $M = 0.1$, $\epsilon=1/33$, $K=0.1$ (top-left), $M = 1$, $\epsilon=1$, $K=\sqrt{0.1}$ (top-right), $M = 1$, $\epsilon=1$, $K=\sqrt{0.5}$ (bottom-left), and $M = 1$, $\epsilon=1$, $K=4$ (bottom-right).}
\label{case4del002}
\end{center}
\end{figure}

\clearpage

\begin{figure}[ht]
	\begin{center}
		\includegraphics[width=12cm, height=8cm]{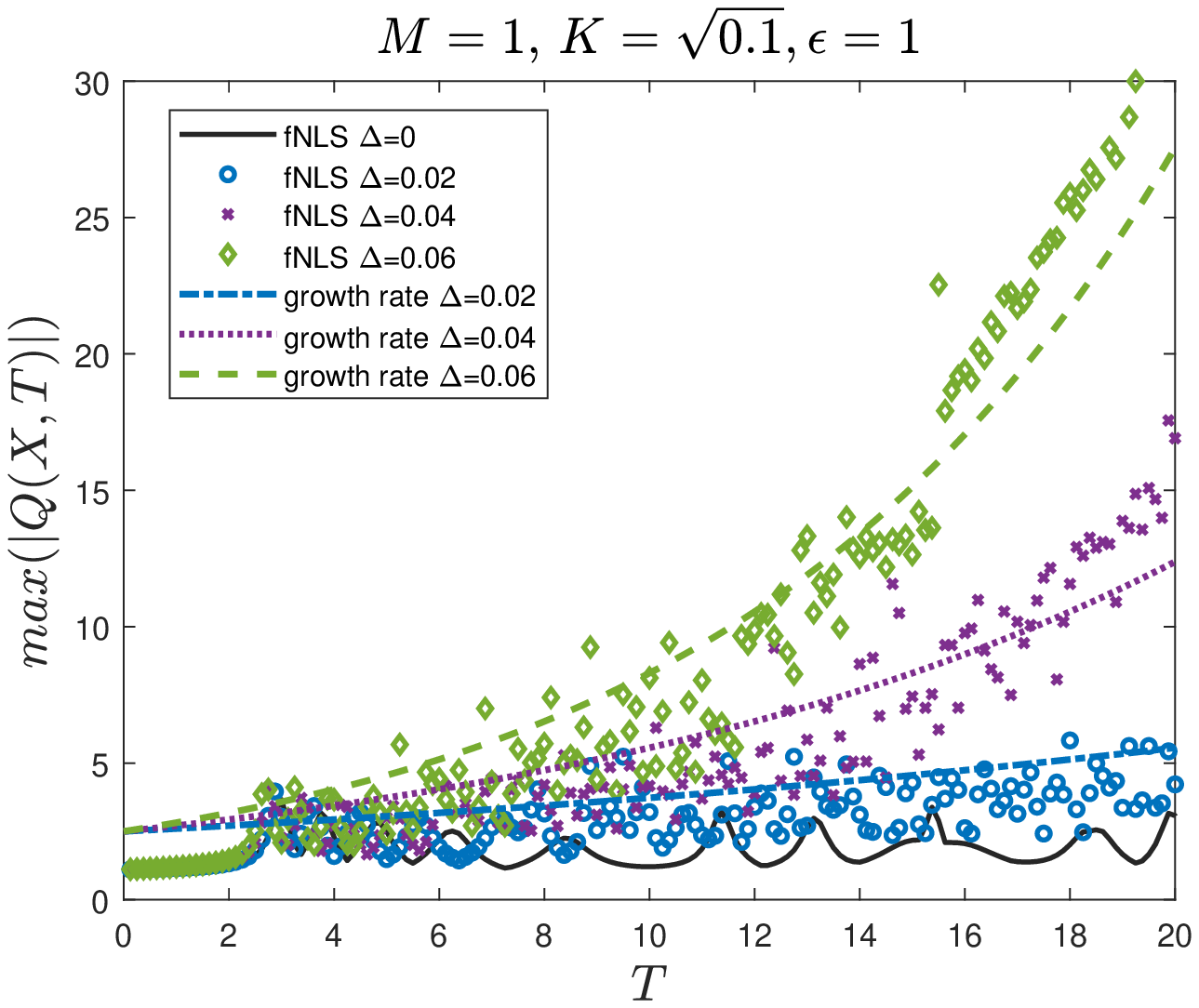}
		\caption{Case 4: The initial condition is (\ref{MIic}).  Maximum of $|Q(X,T)|$ and growth rate $2\Delta$  when $M = 1.0$, $\epsilon=1.0$, $K=\sqrt{0.1}$ for various values of $\Delta$.}
		\label{fig-case4-theory2}
	\end{center}
\end{figure}

\clearpage

\begin{figure}[ht]
	\begin{center}
		\includegraphics[width=12cm, height=8cm]{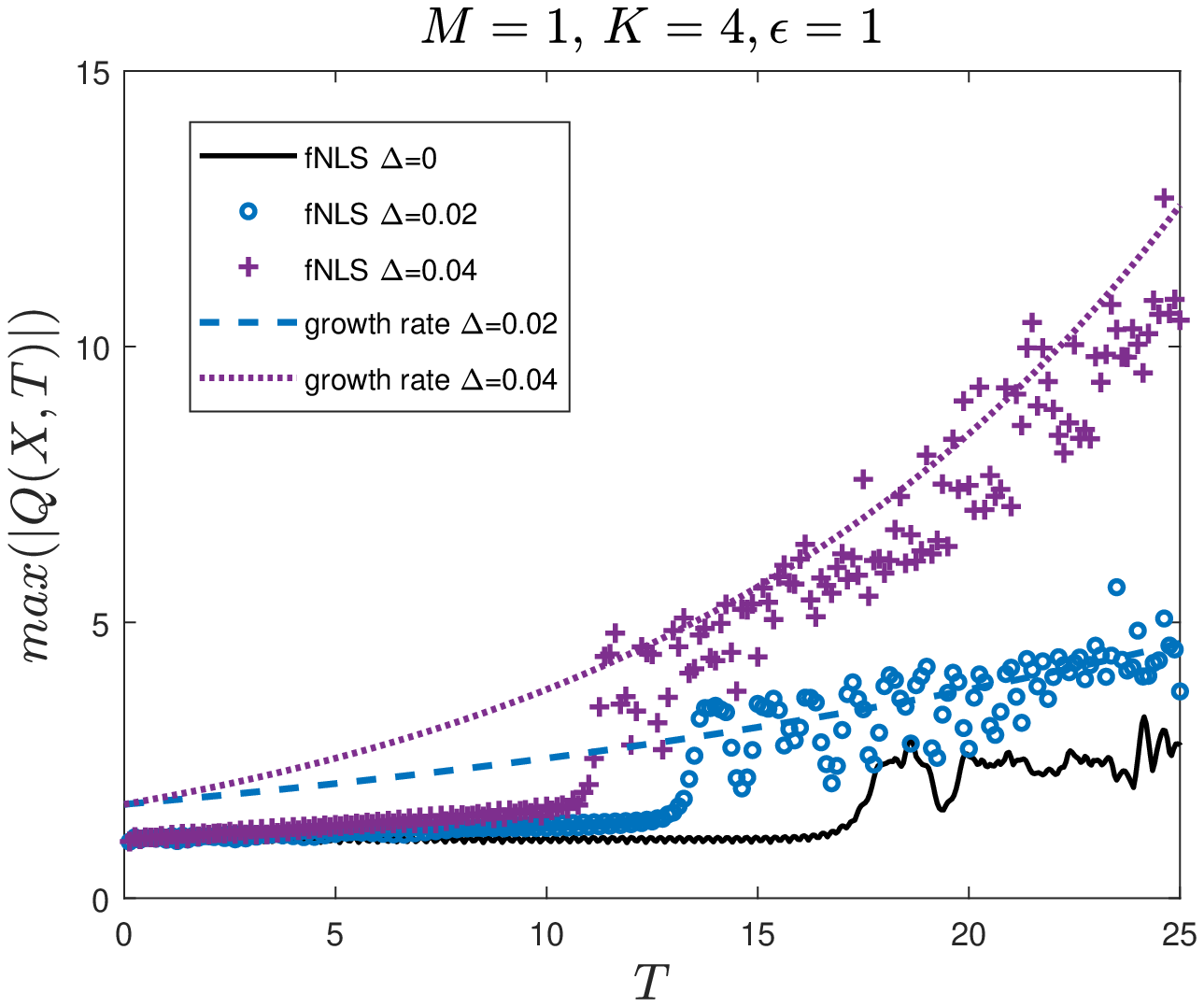}
		\caption{Case 4: The initial condition is (\ref{MIic}). Maximum of $|Q(X,T)|$ and growth rate $2\Delta$ 
		when $M = 1.0$, $\epsilon=1.0$, $K=4.0$ for various values of $\Delta$.}
		\label{fig-case4-theory1}
	\end{center}
\end{figure}

\begin{figure}[ht] 
	\begin{center}
		\begin{tabular}{cc} 
			\includegraphics[height=5cm,width=8cm]{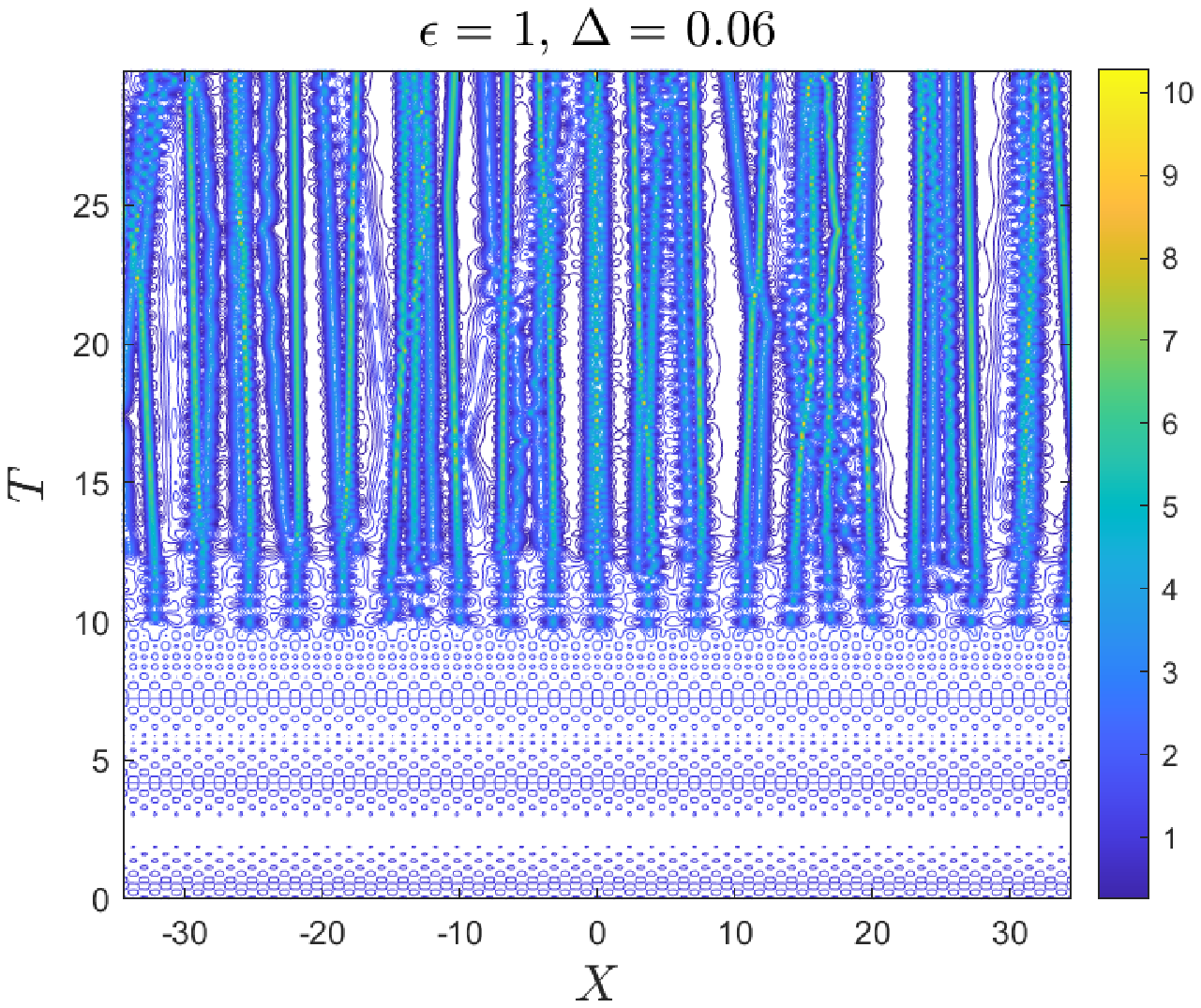}  &
			\includegraphics[height=5cm,width=8cm]{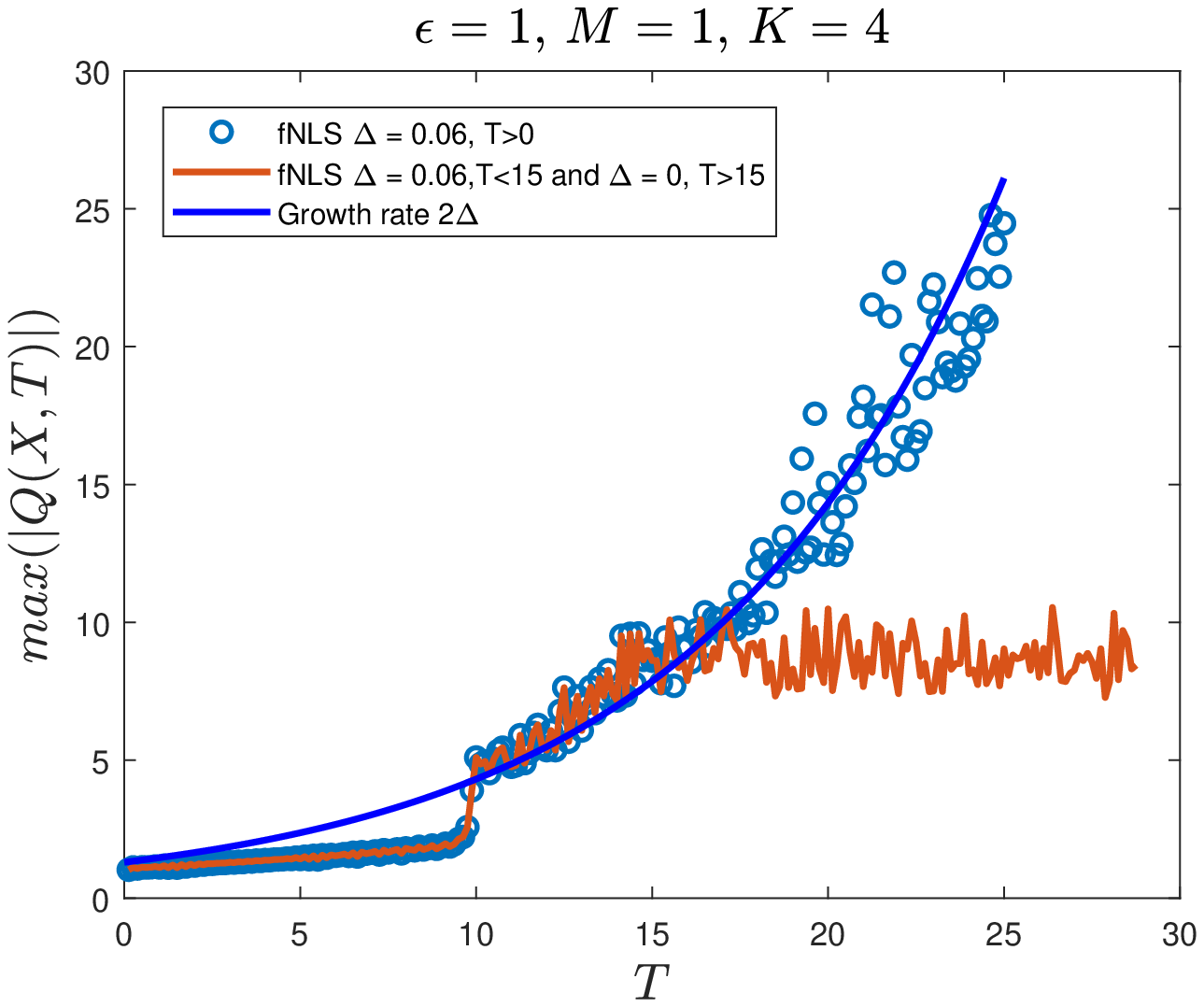}  \\
		\end{tabular}
		\caption{Case 4: The initial condition is (\ref{MIic}). Contour plot and 
		the maximum of $|Q(X,T)|$ with growth rate $2\Delta$  when 
		$M = 1.0$, $\epsilon=1.0$, $K=4.0$ for $\Delta=0.06$. Forcing is turned off when $T>15$}
		\label{case4-turn}
	\end{center}
\end{figure}

\clearpage
\begin{figure}[ht] 
	\begin{center}
		\begin{tabular}{cc} 
			\includegraphics[height=5cm,width=8cm]{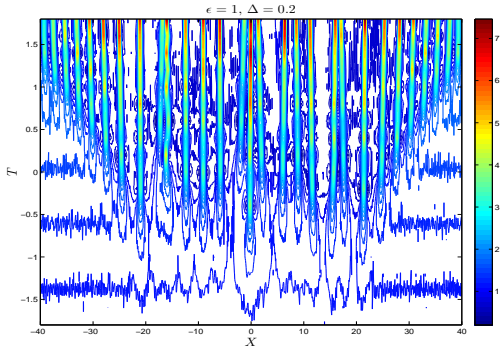}  &
			\includegraphics[height=5cm,width=8cm]{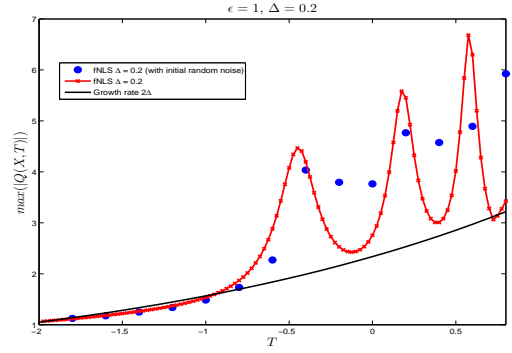} \\
		\end{tabular}
		\caption{Initial random noise, Case 1:  Contour plot of $|Q|$ when 
		$\Delta = 0.2$ (left) and the maximum of $|Q(X,T)|$ with the predicted growth rate $2 \Delta$ (right).}
		\label{case1-rand}
	\end{center}
\end{figure}

\begin{figure}[ht] 
	\begin{center}
		\begin{tabular}{cc} 
			\includegraphics[height=5cm,width=8cm]{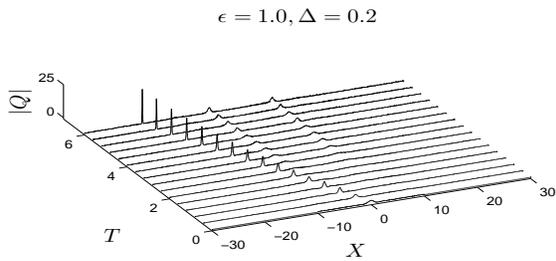} & 
			\includegraphics[height=5cm,width=8cm]{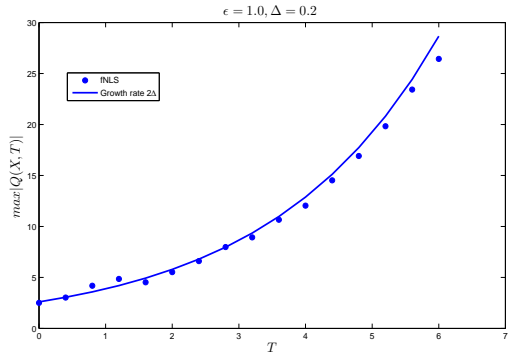} \\
		\end{tabular}
		\caption{Initial random noise, Case 2:  Evolution of $|Q|$ when 
		$\epsilon=1$, $M=2$, $K=-2$ and $\Delta = 0.2$ (left) and the maximum of 
		$|Q(X,T)|$ with the predicted growth rate $2 \Delta$ (right).}
		\label{case2-rand}
	\end{center}
\end{figure}

\begin{figure}[ht] 
	\begin{center}
		\begin{tabular}{cc} 
			\includegraphics[height=5cm,width=8cm]{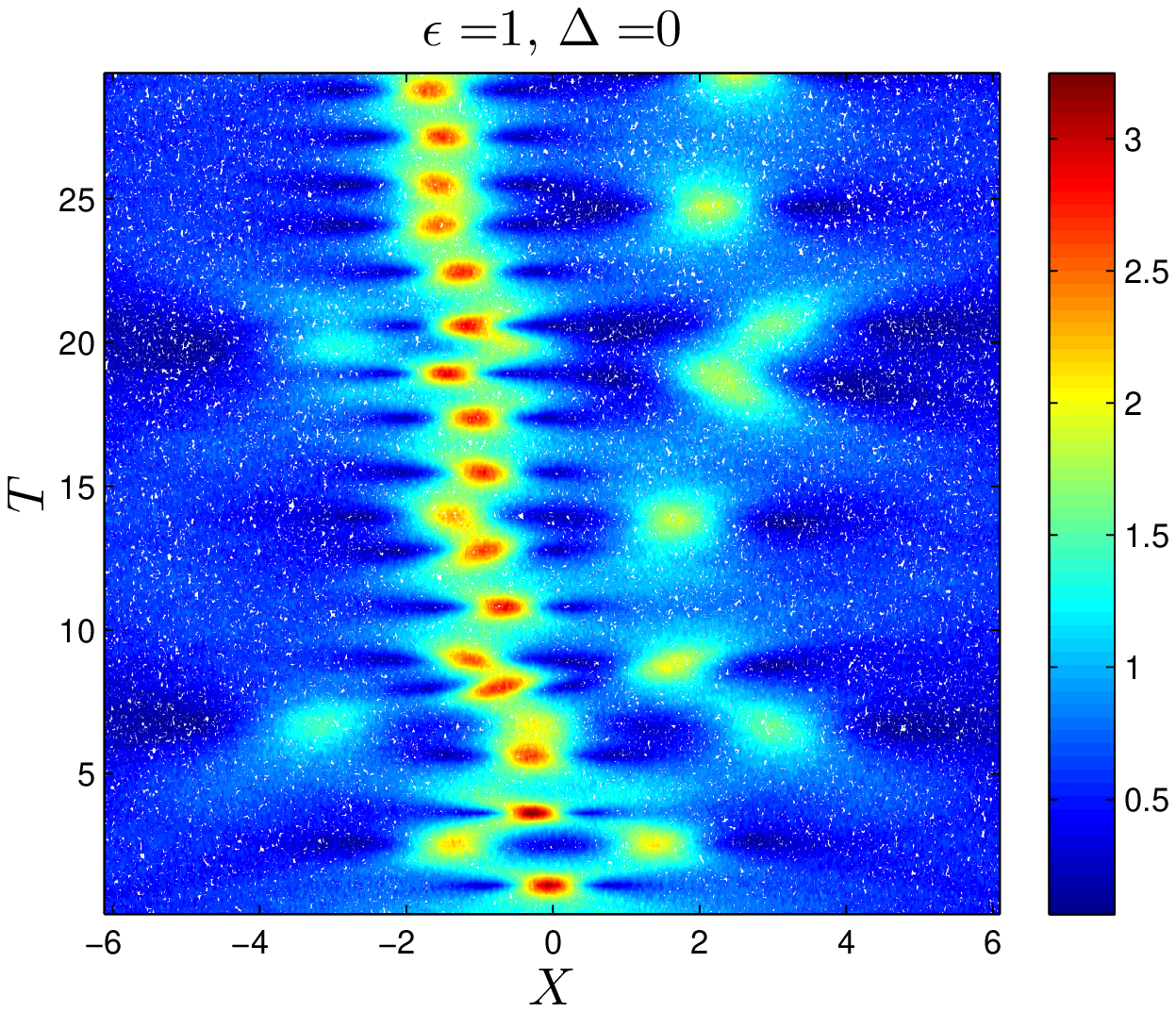}  &
			\includegraphics[height=5cm,width=8cm]{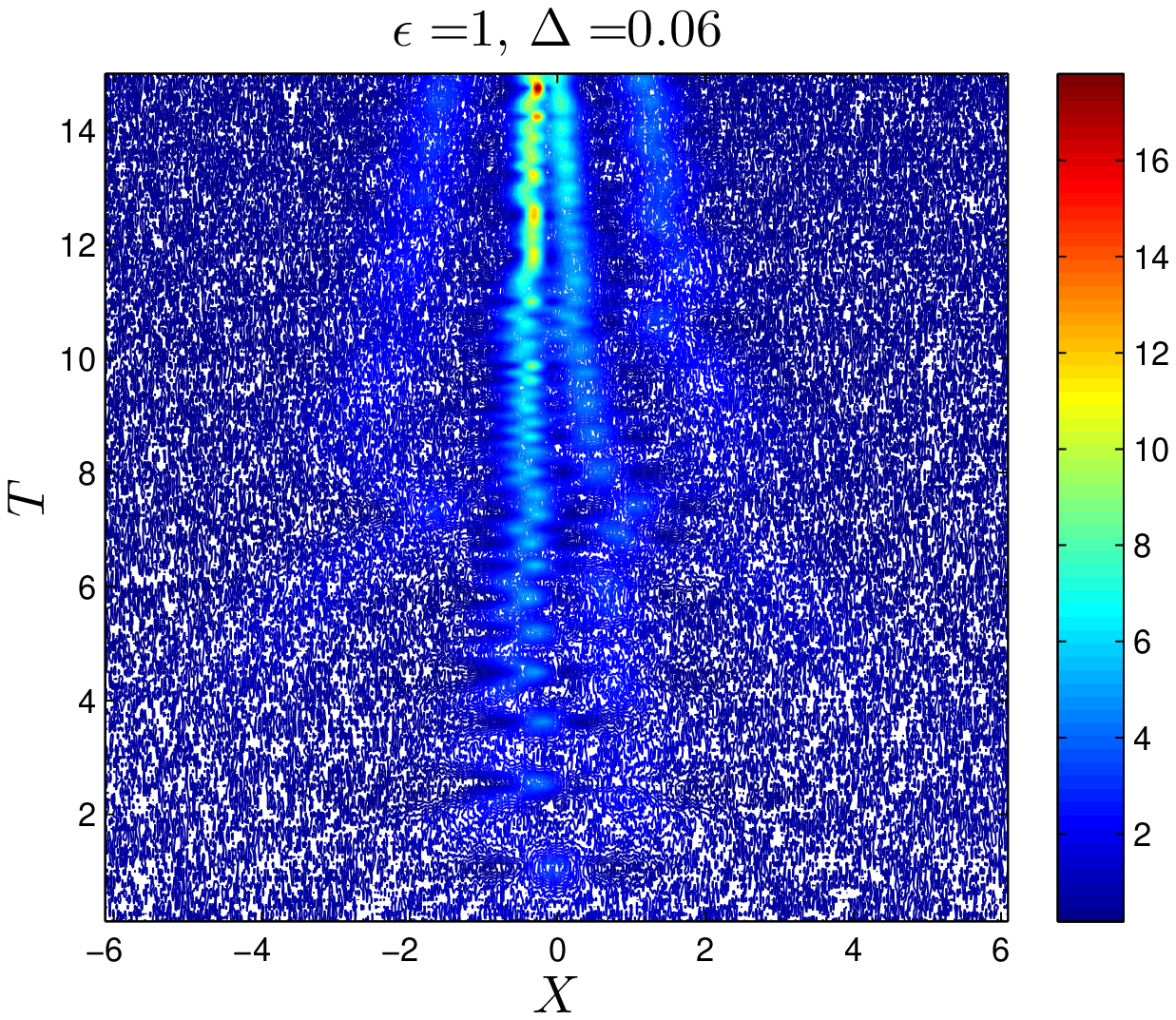} \\
		\end{tabular}
		\caption{Initial random noise, Case 3:  Contour plots of $|Q|$ when 
		$\Delta = 0$ (left) and $\Delta = 0.06$ (right).}
		\label{case3-rand}
	\end{center}
\end{figure}

\begin{figure}[ht] 
	\begin{center}
		\includegraphics[height=7cm,width=10cm]{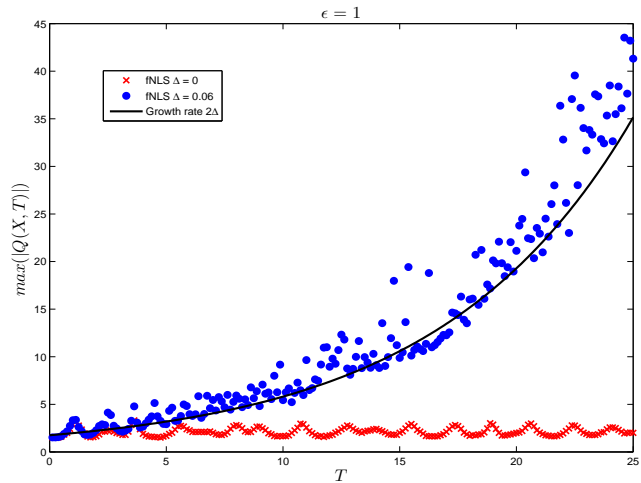} 
		\caption{Initial random noise, Case 3:  The maximum of $|Q(X,T)|$ with the 
		predicted growth rate $2 \Delta$ when $\gamma=0.5$, $\epsilon=1.0$ and 
		$\Delta=0.06$.}
		\label{case3-rand2}
	\end{center}
\end{figure}

\begin{figure}[ht] 
	\begin{center}
		\begin{tabular}{cc} 
			\includegraphics[height=5cm,width=8cm]{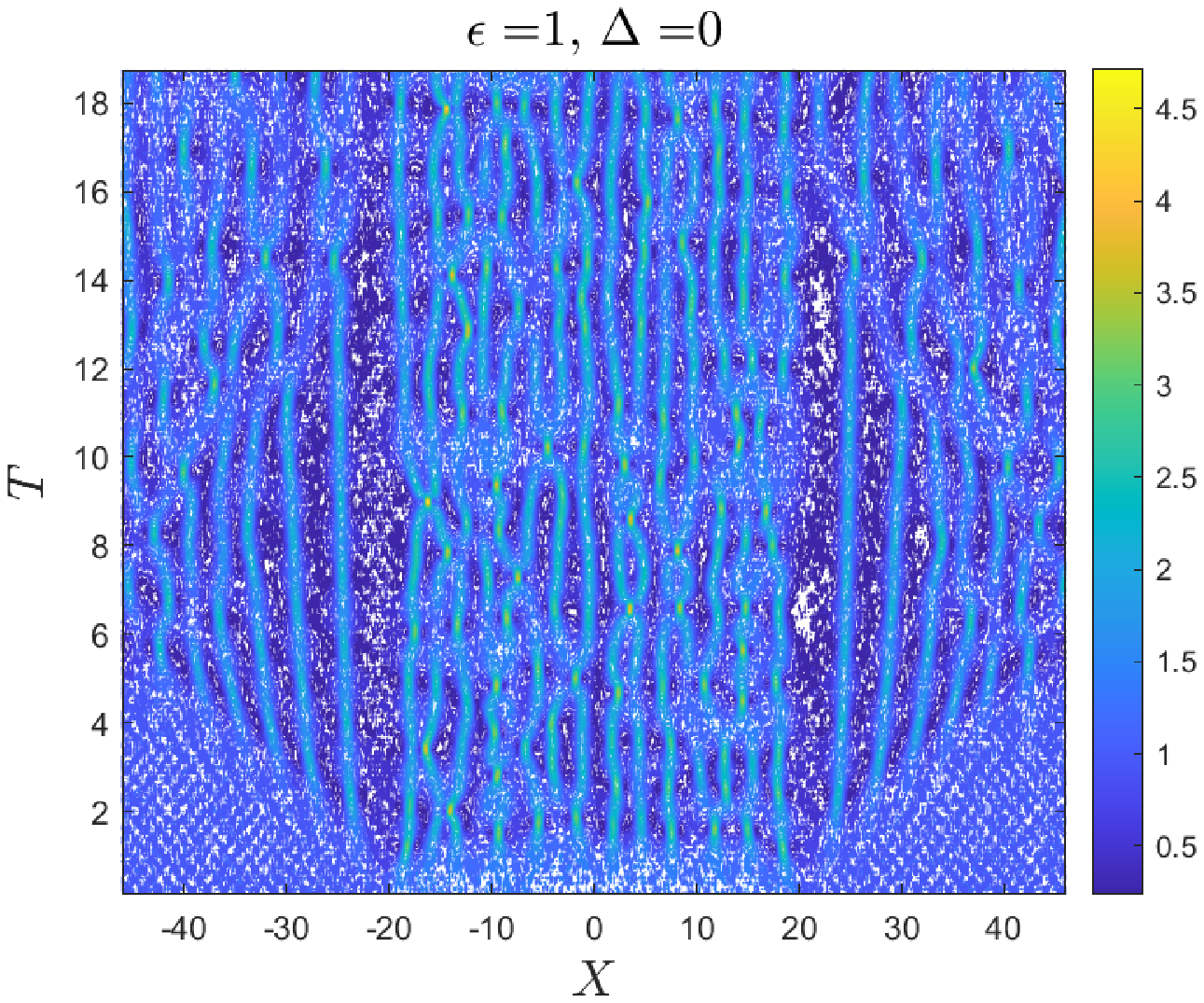}  &
			\includegraphics[height=5cm,width=8cm]{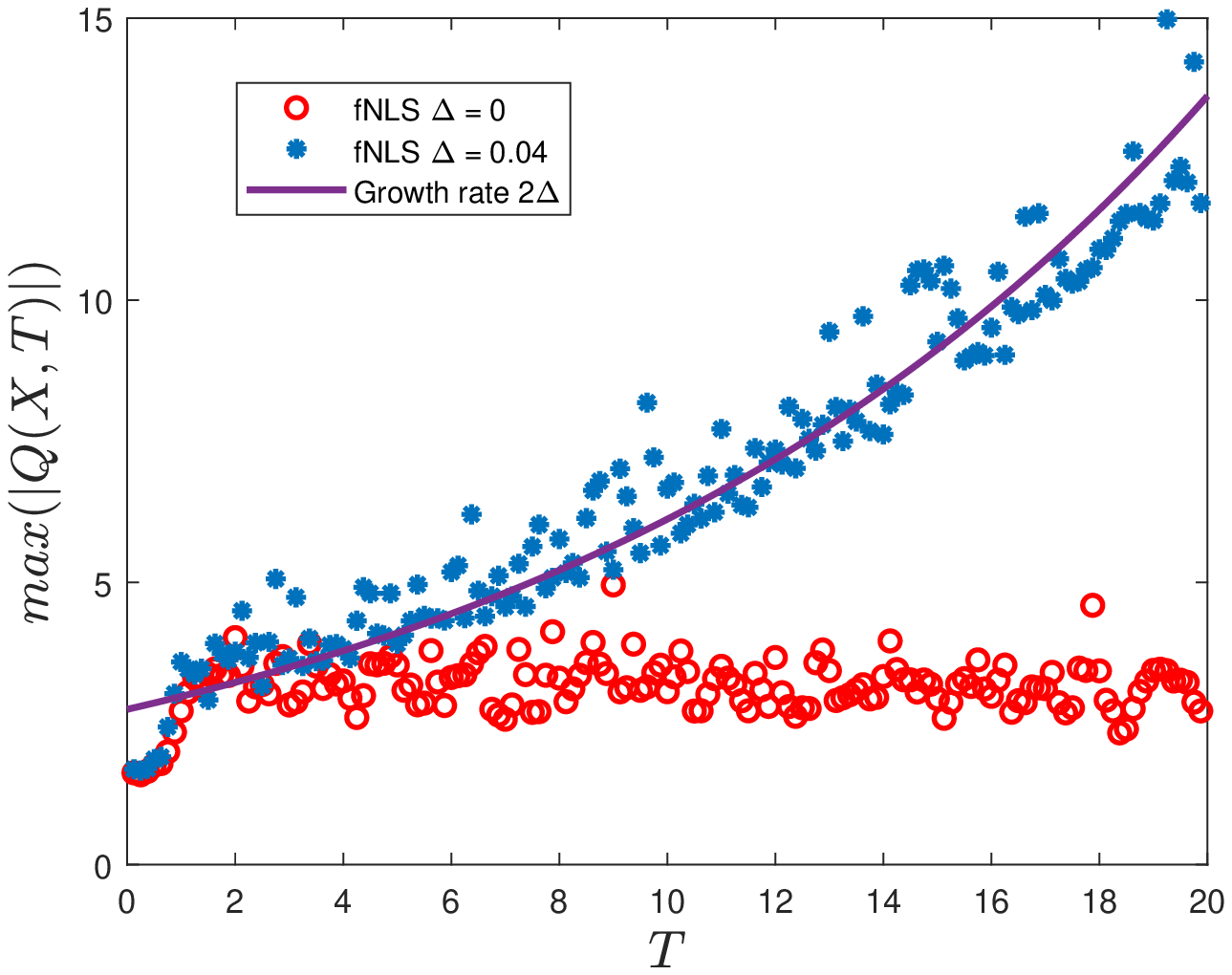} \\
		\end{tabular}
		\caption{Initial random noise,  Case 4:  Contour plots of $|Q|$ when $\Delta = 0$ (left) and the maximum of 
		$|Q(X,T)|$ with the predicted growth rate $2 \Delta$ (right) when $M=1$, $\epsilon=1.0$ and $K=4$.}
		\label{case4-rand}
	\end{center}
\end{figure}

\end{document}